\documentclass[preprintnumbers, floatfix, preprintnumbers, letterpaper, superscriptaddress,nofootinbib]{revtex4}
\usepackage{graphicx}
\usepackage{microtype}
\usepackage{amsmath}
\usepackage{amssymb}
\usepackage{subfigure}
\usepackage{hyperref}
\usepackage{url}
\usepackage{xcolor}
\usepackage{color}
\usepackage{mathrsfs}
\usepackage{calrsfs}
\usepackage{amsfonts}
\usepackage{eufrak}
\usepackage{tabularx}
\usepackage{eucal}
\usepackage{latexsym}
\usepackage{ragged2e}
\usepackage{epsfig}
\usepackage{textcomp}

\usepackage{caption}
\DeclareCaptionJustification{justified}{\leftskip=0pt \rightskip=0pt \parfillskip=0pt plus 1fil}
\definecolor{vividviolet}{rgb}{0.62, 0.0, 1.0}
\definecolor{amaranth}{rgb}{0.9, 0.17, 0.31}
\definecolor{palatinateblue}{rgb}{0.15, 0.23, 0.89}
\definecolor{brightpink}{rgb}{1.0, 0.0, 0.5}
\definecolor{cornflowerblue}{rgb}{0.39, 0.58, 0.93}
\definecolor{deepcarminepink}{rgb}{0.94, 0.19, 0.22}
\definecolor{radicalred}{rgb}{1.0, 0.21, 0.37}

\hypersetup{ linktoc=all,
    colorlinks, linkcolor={palatinateblue},
    citecolor={brightpink}, urlcolor={amaranth}
}

\graphicspath{{Images/}}

\renewcommand{\d}[1]{\ensuremath{\operatorname{d}\!{#1}}}

\graphicspath{{Images/}}
\renewcommand{\d}[1]{\ensuremath{\operatorname{d}\!{#1}}}
\makeatletter
\def\@fnsymbol#1{\ensuremath{\ifcase#1\or \ddagger \or  $\textleaf$  \or \dagger
\else\@ctrerr\fi}}%

\makeatother

\def\sideremark#1{\ifvmode\leavevmode\fi\vadjust{\vbox to0pt{\vss
 \hbox to 0pt{\hskip\hsize\hskip1em
 \vbox{\hsize1.3cm\tiny\raggedright\pretolerance10000
 \noindent #1\hfill}\hss}\vbox to8pt{\vfil}\vss}}}%
\def\beq{\begin{equation}}
\def\eeq{\end{equation}}

\begin{document}

\title{Neutrino oscillations inside ultrarelativistic matter: on the role of the Tolman VI spacetime}

\author{Daniele \surname{Gregoris}}
\email{danielegregoris@libero.it}
\affiliation{Centre For Cosmology and Science Popularization (CCSP), SGT University, Gurugram, Delhi-NCR, Haryana- 122505, India}

\begin{abstract}
While both neutrino propagation in photons and ultrarelativistic plasmas and neutrino oscillations in curved spacetimes have been studied extensively, their combination in a self-gravitating geometry appears not to have been considered previously. In this paper, we will investigate neutrino oscillations inside a static spherically symmetric spacetime, derived self-consistently from Einstein's equations, supported by  ultrarelativistic matter.  By relying on the geodesic motion and geometrical symmetries of the configuration, we will compute analytically  the neutrino eikonal and spinoptics phase oscillations. Next, we will relate  resonance phenomena to the physical interpretation of the background material inside which neutrinos travel for  disentangling gravitational versus electroweak effects.  We will inspect the effects  of the pure gravitational and pure matter fields, and of the temperature in the neutrino oscillations; we will explain why our results are not coordinate artifacts, and we will certify our heuristic method by comparing with the asymptotic behaviours predicted by thermal field theory approaches. We will raise awareness on the relevance of the spinoptics phase shift  entering  the flavor conversion probability, should the neutrinos hit a curvature singularity.  Theoretically, we will scrutinize  the separate  roles of geometrical symmetries versus physical features of the configuration, while  astrophysically we will qualitatively discuss neutrino oscillations inside stars and in the Cosmic Microwave Background at different temperature and density conditions.  Our research puts forward, in the context of the Tolman VI metric, an observational method for identifying naked singularities in our Universe: neutrinos would simultaneously be physically overlapped, in flavor-equilibrium and quantum-mechanically scrambled.
\end{abstract}

\maketitle

\section{Introduction}
\label{ss1}

Neutrinos impact the evolution of both large/small scale and high/low density gravitational configurations: consequently, they are investigated  within both the cosmological and stellar  realms. In the former context, neutrinos contribute to the energy budget  of the universe affecting its evolution all along the cosmic history from early times, at which leptogenesis occurred, up to the era of formation of bound astrophysical structures: we refer to \cite{dolgov} for a review of  observational evidences and theoretical models accounting for these phenomena. Regarding the latter scenario instead, neutrinos may not only trigger the collapse of massive stars leading to supernovae explosions, but also shape the mergers of neutron stars;  signatures of the presence of neutrinos could be  observed  in the nucleosyntheis and formation of heavy elements \cite{stars}. At intermediate regimes, self-gravitating particles with energy scales of neutrinos have been proposed to constitute the dark matter filling our galaxy and sustaining a dense  dark core whose characteristics could explain astrophysical data customarily attributed to a black hole \cite{carlos}. Moreover, measurements of neutrino oscillations can be utilized for inferring the distribution of dark matter halos around compact objects \cite{infDM}.

A specific phenomenon theoretically predicted by Pontecorvo  deals with the oscillations of neutrino flavors \cite{pontecorvo1,pontecorvo2}.  This reasoning proceeded by analogy, and the core intuition was to postulate that if electrically neutral mesons mix because their flavors and masses are misaligned \cite{gellmann}, then neutrinos --  fermions, but also neutral -- should in principle behave the same unless a symmetry forbids it. The supporting model builds upon Majorana's idea that neutral fermions can be self-conjugate \cite{majorana}.  Pontecorvo's proposal did not attract much attention at the beginning because it requires to abandon the two-component Weyl neutrino theory according to which neutrinos have exactly zero mass\footnote{It should be noted though that dissipative effects can trigger oscillations also for initially massless neutrinos \cite{benattiA}.}; Pontecorvo's theoretical prediction could be observationally confirmed only much later thanks to the Super-Kamiokande Collaboration which has tracked  atmospheric neutrino oscillations \cite{exp1} and further  by measuring the solar neutrino flavor conversion \cite{exp2}. Eventually, with the discovery of the third family of neutrinos, available experiments cannot discriminate  between Dirac and Majorana models \cite{bilenky}. Here, we should recall that the definitions of Weyl, Dirac and Majorana fermions rely on the geometry of the proper Lorentz group and not on their behaviors under discrete symmetries (such as parity and charge conjugation) \cite{american}: Weyl's are massless particles with fixed helicity, Dirac's is a particle that is distinct from its antiparticle, and Majorana's is a particle which is its own antiparticle.  Specifically, the future planned experiment JUNO is expected to establish the mass hierarchy of the neutrinos \cite{juno}.

Although neutrinos are genuinely wavepackets in nature, whose modeling requires a quantum dynamical treatment, they can still be \lq\lq heuristically" described as point particles following geodesic paths of a given curved spacetime manifold which account for the background spatial region they cross \cite{heuristic}. This approach, also known as \lq\lq eikonal approximation", boils down to a geometric optics argument as that invoked in cosmology for  tracking the light ray signals arriving from supernovae explosions which are at the core of the  luminosity distance relation datasets exploited for testing cosmological models;  its applicability to the description of the motion of neutrinos is supported also by an analogy with the double-slit experiment \cite{slit}. This picture of traveling neutrinos has been widely adopted with applications to the estimate of their oscillation probability within a number of  gravitational systems given by curved spacetimes --such as black holes, stars, gravitational waves and homogeneous and isotropic universe --in the realm of gravitational relativistic theories \cite{visinelli,burgard,app}; independently of the technicalities and specific outcomes qualifying the analyses in each of these spacetimes, the take-at-home message is that neutrino oscillations can be induced not only by electroweak interactions, but also by the presence of a gravitational field: this insight offers the possibility of computing their conversion probability  making abstraction from the formulations of electroweak theories. In this framework,  the physical idea is that  the neutrinos do not scatter against a  medium, but they interact indirectly with it via the Christoffel symbols of the geodesic equation arising for the spacetime supported by such a specific  material \cite{miotolman,indi}.

A refined description of the neutrino motion, taking into account the spin of the particle, has been recently developed in \cite{frolovA} and subsequently applied in the cases of  Schwarzschild \cite{frolovB} and Kerr \cite{frolovC} black holes. In this paper, we will explicitly confirm that the spinoptics contribution to the phase oscillation and conversion probability    for neutrinos traveling inside the Tolman VI metric \cite{tov1} is indeed subdominant with respect to that from the geometrical optics but for the spatial region containing the central naked curvature singularity arising in this spacetime. Spinoptics constitutes an elegant mathematical formalism which allows us to overcome the complex mathematics of the $4 \times 4$ spinor matrices by simply treating the spin as a  vector; physically,  the eikonal phase relates to the translational motion of the neutrinos, while the spinoptics phase to their rotational or frame motion. On the other hand,  Tolman VI metric describes a spacetime filled with self-gravitating particles in gravitational equilibrium and thermodynamically accounted for at the macroscopic level by a one-parameter barotropic equation of state; by suitably adjusting the numerical value of such a parameter, the equation of state reduces to that of ultrarelativistic particles possibly constituting radiation or electron-positron $e^-e^+$ plasma. Therefore, this assumption for the region inside which neutrinos live will allow us to describe the phenomenon of neutrino oscillations at different temperature and density regimes,  as those we have reviewed at the beginning of this paper (early and late time universe, and interiors of stars), in a unified manner. Although our considered spherically symmetric gravitational and matter fields  cannot generate any net asymmetry between neutrino helicities,  it should nevertheless be appreciated that our results will not only complement literature ones based on  separate quantum field theory approaches for the description of neutrino oscillations inside radiation and photon media, but they will actually be more general thanks to the possibility of tuning  the free parameter in the Tolman VI metric's equation of state.

In this paper, first of all, we will  compute the conversion probability for neutrinos traveling inside the Tolman VI metric essentially following the gravitational physics perspective we have  outlined above; we will then corroborate the correctness of our method by checking that the asymptotic limits  at low and high temperature match literature results based on completely different arguments: for a low temperature environment neutrinos would behave as if they were traveling inside an empty flat spacetime, while within a hot medium the eikonal neutrino flavor survival probability tends to one but the spinoptics contribution locks the system into a state of flavor equilibrium.  Next, relying on the theory of electroweak interactions,  we will assess the possible occurrence of Mikheyev-Smirnov-Wolfenstein (MSW in short) resonances \cite{MSW}; in this latter context, the physical interpretation of the nature of the particles realizing the ultrarelativistic equation of state for the material filling the spacetime becomes crucial: for example, if we regard it as a photon gas, then no resonance at all can occur  because of the vanishing of the necessary potential. Therefore, these two portions of our manuscript, on the oscillations and resonances, will exploit the geometrical symmetries and the physical features of the Tolman  VI metric separately.

Our paper is organized as follows. In Sect.\ref{ss2}, we will present the spacetime metric inside which neutrinos travel and review some of its basic properties, like its temperature gradient via Tolman's law and its geodesic equations of motion. In Sects.\ref{ss3}-\ref{ss4}, we will firstly compute the neutrino phase oscillation in the eikonal (or geometric optics) approximation and then adopting the spinoptics formalism; here, we will  carefully inspect the roles of the geometrical symmetries of the underlying spacetime, and neatly relate the phase shift and conversion probability to the temperature of the background matter density and to the strength of the gravitational field separately; we will also establish the meaning of the free model parameter entering the Tolman VI metric in determining the locations of the maxima/minima of the conversion probability, and we will constrain the range of applicability of our framework by computing the coherence length of the neutrino wavepackets. In Sect.\ref{sectprob}, we will exhibit our result for the conversion probability and show that our geometrical procedure correctly captures the qualitative behaviors of the oscillations at both low at high temperature regimes without the need of carrying out quantum field theory resummations. On the other hand, in Sect.\ref{ss5}, we will address the thermodynamical interpretation of the equation of state of the material inside which neutrinos travel and oscillate, and possibly rule out (as for a photon gas) or quantify (as for an $e^-e^+$ plasma either superimposed as a test field or self-gravitating) resonance phenomena. We will conclude in Sect.\ref{ss6} by summarizing our methods and results,  and by providing an outlook for future research. We emphasize since now that all our considerations will be based on exact analytical closed-form computations.

\section{Review of the Tolman VI metric}
\label{ss2}

Our spacetime metric of interest reads as\footnote{In this paper, if not otherwise stated, we will adopt units such that $8\pi G=1=c^2$.}
\beq
\label{metric}
\d s^2=g_{\mu\nu} \d x^\mu \d x^\nu=-r \d t^2 + a \d r^2 +r^2(\d\theta^2 + \sin^2 \theta \d\phi^2)\,,
\eeq
where $a$ is a (constant) free model parameter. This spacetime, if interpreted as a solution of the Einstein field equations of General Relativity $G_{\mu\nu}=T_{\mu\nu}$, is filled by a perfect fluid with stress-energy tensor $T^{\mu}{}_{\nu}={\rm diag}[-\rho,\, p_r,\, p_t,\, p_t]$ whose energy density and pressure are given by
\beq
\label{einstein}
\rho=\frac{a-1}{ar^2}\,, \qquad p_r=\frac{2-a}{ar^2}\,, \qquad p_t=\frac{1}{4a r^2}\,.
\eeq
A naked curvature singularity located in $r=0$ should be noted. Weak energy condition $\rho \geq 0$ restricts $a \geq 1$ or $a<0$, while spacetime causality  rules out the latter possibility; the boundary choice $a=1$, which realizes an Euclidean spatial element with no geometrical deficit, should also be excluded because it delivers a non-zero material pressure with vanishing energy density. For the specific choice $a=7/4$, the pressure becomes isotropic and the stress-energy tensor tracefree, implying that the spacetime is supported by a  gas of ultrarelativistic particles (possibly photons) obeying the equation of state $p=\frac{\rho}{3}$: in fact, in General Relativity also the photons carry a \lq\lq weight" generating spacetime curvature \cite{tolmanold}.  Eq.(\ref{metric}) constitutes a subcase of the so-called Tolman VI metric \cite{tov1} whose geodesics augmented with scattering and frictional effects have been investigated in\footnote{Note that we are writing the spacetime metric is a slightly different form rescaling the time coordinate.} \cite{miotolman}.

The  curvature of our assumed spacetime with $a=7/4$ could be generated by an effective averaged electromagnetic radiation: the expression of the stress-energy tensor arises  from coarse-graining a bath of electromagnetic waves over spacetime regions large enough compared to the electromagnetic wavelength and in such a manner that the energy distribution is isotropic and can be treated as a perfect fluid (this operation constitutes an ensemble average over many incoherent modes). What is eventually obtained  is the so-called  Isaacson's high-frequency stress-energy  tensor holding under the following assumptions \cite{ISA}: {\it (i)} the photons' mean free path is much smaller than the spacetime curvature scale; {\it (ii)} the energy distribution is isotropic in the local rest frame; {\it (iii)} coherence lengths are microscopic.

According to Tolman's law \cite{tolmaneq}, which determines the constancy of the redshifted temperature across the system, the radial decay of the locally measured temperature for a configuration in gravitational equilibrium is $T_\infty=T \sqrt{-g_{tt}}=T \sqrt{r}$, where in the last step we have enforced our (\ref{metric}); in what follows it is understood that $T\equiv T(r)$ and a subscript infinity indicates that the corresponding quantity is measured at spatial infinity. Tolman's law for the temperature in our considered spacetime (\ref{metric}) then implies the  Stefan-Boltzmann law $\rho \propto T^4$ \cite[p.186]{landau}.  It should be noted that the scaling behavior as a fourth power in the latter is independent of the numerical value prescribed to $a$ (the Stefan-Boltzmann's proportionality constant instead depends on it) because it follows from the redshift factor and the equilibrium conditions set by the Einstein field equations and not from the microphysics of the equation of state determining instead the spatial geometry. In other words, in the Tolman VI metric, the Stefan-Boltzmann law stems as a consequence of the scaling properties of the spacetime (that is a self-similar solution for which every quantity  depends only on the radial coordinate $r$ preserving their inter-relationships  like the equation of state\footnote{We can note that for any generic value of the parameter $a$, the ratios between energy density, radial and tangential pressure are independent of the surface on which they are computed.} \cite{carr}), and not from having a radiation fluid. 

We will now review some basic facts about the geodesic motion in the Tolman VI metric; we will need them for computing the phase shift and for inferring the physical effects generating the conversion probability between different neutrino flavors. In what follows, $\lambda$ will denote the affine parameter along an either timelike or null geodesics. First of all, since the spacetime is static, the energy $E$ of a particle moving along a geodesic path is conserved:
\beq
U^t=\frac{\d t }{\d\lambda}=\frac{E}{r}\,.
\eeq
Moreover, since the spacetime is spherically symmetric, we can choose the geodesic motion to take place on the equatorial plane $\theta=\frac{\pi}{2}$ with $U^\theta=\frac{\d\theta}{\d\lambda}=0$, and the conservation of angular momentum $L$ holds:
\beq
U^\phi=\frac{\d\phi }{\d\lambda}=\frac{L}{r^2}\,.
\eeq
Lastly, the normalization condition $U^\mu U_\mu=\epsilon$ for the velocity four-vector, where $\epsilon=0,\, -1$ for null and timelike trajectories respectively, allows us to obtain the radial component of the vector tangent to the geodesic path as
\beq
\label{radial}
U^r=\frac{\d r}{\d \lambda}=\pm \sqrt{\frac{1}{a}\left(\epsilon +\frac{E^2}{r}-\frac{L^2}{r^2}\right)}\,,
\eeq
where the plus/minus sign corresponds to an outgoing/ingoing trajectory respectively.
Therefore, during their motion,  the neutrinos experience the effective gravitational potential
\beq
V_{\rm eff}(r)= \frac{L^2}{r}-\epsilon r \,,
\eeq
which permits also  a (closed) circular timelike orbit with radial coordinate such that\footnote{The double sign is such that $r$ shall be positive.} $\frac{\d V_{\rm eff}}{\d r} \Big|_{r=\pm L}=0$.  We note that the free parameter $a$ qualifying the type of material (\ref{einstein}) enters only the radial component of the four-velocity. We  remark also that, in general, the values of (conserved) energy $E$ and angular momentum $L$ are different for particles following either  null or timelike paths: we nevertheless denote them the same because they are  taken to match each other in the specific  physical application to the phase change of neutrino oscillations \cite{heuristic,visinelli,burgard,app}; in fact, it is customarily assumed that the neutrino mass eigenstates correspond to their energy eigenstates with a common energy $E$ \cite{stodo,bilenky2}.

\section{Neutrino oscillations in the geometric optics approximation}
\label{ss3}

Under the WKB approximation and in the eikonal limit, the  phase change of neutrino flavor oscillations must be computed from \cite{heuristic}
\beq
\Phi_{kj}=\Phi_k - \Phi_j \,, \qquad \Phi_k\equiv \Phi_k^{\rm eik}=\int_{r_A}^{r_B}P_{\mu (k)}\frac{K^\mu}{K^r}\d r\,,
\eeq
where the subscripts $k$ and $j$ identify the $k$ and $j$ species of neutrinos which are emitted at the spacetime location $r_A$ and detected in $r_B$, while $P_{\mu (k)}=m_k U_\mu$ with $U_\mu$ and $K_\mu$   timelike and null geodesic  velocity four-vectors. We note that the phase change, in a diagonal spacetime metric like ours,  is not sensitive to whether the neutrinos travel outwardly or inwardly because  the product between  radial components of momentum and null four-velocity  washes away the double sign possibility in (\ref{radial}). Regarding the mathematical formula above for the phase change, we should keep in mind that neutrino detectors measure only distances and not time intervals \cite{stodolsky,lipkin}. We use the word \lq\lq phase"  because the solution of the Dirac equation $[i\gamma^\mu(\partial_\mu + \Gamma_\mu)+m]\psi=0$ for the mass eigenstate is given by $T^\dagger |\nu_k \rangle=e^{i \Phi_k}  |\nu_k \rangle$; here, $\Gamma_\mu$ and $T^\dagger$ denote the spinor connection and the spacetime evolution operator, respectively. We refer to \cite[Sects-2-3]{visinelli} and  \cite[App.A]{referee2} for self-contained reviews about the covariant treatment of the Dirac equation in a generic curved spacetime. We just emphasize that relevant assumptions for adopting this formalism are met: {\it (i)} our assumed gravitational theory, General Relativity, comes without torsion (meaning that the Christoffel symbols governing the geodesic paths reduce to the Levi-Civita connection); {\it (ii)} the spacetime through which neutrinos propagate is non-rotating.

In the configuration of our interest (\ref{metric}), we  explicitly obtain: 
\begin{eqnarray}
\Phi_k^{\rm eik} &=& \sqrt{a} \int^{r_B}_{r_A}\left[\sqrt{-m_k^2+\frac{E^2}{r}-\frac{L^2}{r^2}}-\sqrt{\frac{E^2}{r}-\frac{L^2}{r^2}} \right]\d r \\
&=& \sqrt{a} \bigg\{ \sqrt{E^2 r- L^2-(m_kr)^2}- \frac{E^2}{2m_k} \arctan\left(\frac{E^2-2m_k^2r}{2m_k\sqrt{E^2 r- L^2-(m_kr)^2}}\right)\\
&&\quad +L \ln \bigg| \frac{E^2 r- 2L^2+2L \sqrt{E^2r-L^2-(m_kr)^2}}{r}\bigg|   + 2\left[ L \arctan \left(\frac{\sqrt{E^2 r-L^2}}{L} \right)-\sqrt{E^2 r-L^2}\right]\bigg \} \Bigg|^{r_B}_{r_A}\,.
\end{eqnarray}
Should the neutrinos be traveling along a purely radial path, the previous result simplifies to:
\beq 
\Phi_k^{\rm eik}=\sqrt{a}\left[ \sqrt{(E^2 -m_k^2r)r}-\frac{E^2}{2m_k}\arctan\left(\frac{E^2-2m_k^2r}{2m_k\sqrt{(E^2 -m_k^2r)r}} \right)-2E\sqrt{r}\right] \Bigg|^{r_B}_{r_A}\,.
\eeq
Then, for ultrarelativistic neutrinos  for which $m_k \ll E$, we find\footnote{According to the equation below, neutrinos can reach   the center of the configuration only following radial paths. In light of Tolman's law, that is the region at highest temperature.}:
\begin{eqnarray}
\label{finP0}
\Phi_{kj}^{\rm eik} &\approx& \frac{ \sqrt{a} }{2E} \int^{r_B}_{r_A} \left( \Delta m^2(r-b^2)^{-1/2} r -\frac{\Delta m^4 r^3}{4E^2}(r-b^2)^{-3/2}\right) \\
\label{finP}
&=& \frac{ \sqrt{a} }{E}\left(\frac{(r+2b^2) \sqrt{r-b^2}\Delta m^2}{3}-\frac{(r^3+2(br)^2+8b^4r-16b^6)\Delta m^4}{20E^2\sqrt{r-b^2}} \right) \Bigg|^{r_B}_{r_A} \,,
\end{eqnarray}
where we have introduced the impact parameter $b:=L/E$, and the mass variations $\Delta m^2:=m^2_k-m^2_j$ and $\Delta m^4:=m^4_k-m^4_j$.

Thus, we have obtained, consistently with the expectation from general arguments \cite[Eq.(32)]{heuristic},  that  the leading order  term is modified by simply replacing the crossed flat spacetime  distance with an effective curved one:
\beq
\label{Reff}
\Phi_{kj}^{\rm eik} \approx  \frac{ \Delta m^2 R_{\rm eff} }{2E} \qquad {\rm with} \qquad R_{\rm eff}= \int \sqrt{-g_{rr} \cdot g_{tt}} \d r \,.
\eeq
Let us now put forward three interpretations of the eikonal  phase shift grounded on the strength of the pure matter field, on the temperature gradient, and on the strength of the pure gravitational field, separately. In fact, our obtained eikonal phase shift relates to the background energy density (\ref{einstein}) as:
\begin{eqnarray}
\label{eikrho}
&&\Phi_{kj}^{\rm eik}\approx\frac{ 1 }{E (a^3\rho^5)^{1/4}\sqrt{\sqrt{a-1}-b^2\sqrt{a\rho}}}\Bigg(\frac{(a-1+b^2\sqrt{(a-1)a\rho}-2b^4a\rho)\sqrt{a\rho}\Delta m^2}{3}\\
&& \qquad \qquad \qquad \qquad \qquad \qquad \qquad \qquad \qquad  +\frac{16b^6(a\rho)^{3/2}-8b^4\sqrt{a-1}a\rho-2(a-1)b^2\sqrt{a\rho}-(a-1)^{3/2}}{20E^2} \Bigg) \Bigg|^{\rho_B}_{\rho_A} \,.
\end{eqnarray}
On the other hand, keeping in mind Tolman's law for the temperature gradient, the eikonal phase shift can be recast as
\beq
\label{eikT}
\Phi_{kj}^{\rm eik}\approx \frac{ \sqrt{a} }{E}\left(\frac{[T_\infty^2+2(bT)^2] \sqrt{T_\infty^2-(bT)^2}\Delta m^2}{3T^3}-\frac{[T_\infty^6+2(bTT_\infty^2)^2+8(b^2T^2T_\infty)^2-16(bT)^6]\Delta m^4}{20E^2\sqrt{T_\infty^2-(bT)^2}T^5} \right) \Bigg|_{T_A}^{T_B}\,.
\eeq
Finally, the eikonal phase shift can  be related also to the strength of the pure gravitational field (the latter quantified by the conformal curvature to be contrasted against the Ricci component accounting for the pure matter field whose effects we have studied in (\ref{eikrho})) as 
\begin{eqnarray}
&&\Phi_{kj}^{\rm eik}\approx \frac{1}{2E}\sqrt{\frac{a}{2(-6a \Psi)^{3/2}(\sqrt{4a-1}-2b^2\sqrt{-6a\Psi})}}\Bigg(\frac{(\sqrt{4a-1}+4b^2\sqrt{-6a\Psi})(\sqrt{4a-1}-2b^2\sqrt{-6a\Psi})\Delta m^2}{3}  \\
&& \qquad \qquad \qquad  \qquad  -\frac{[(4a-1)^{3/2}+8b^2(4a-1)\sqrt{-6a\Psi}+192ab^4\sqrt{4a-1}\Psi-128b^6(-6a\Psi)^{3/2}]\Delta m^4}{40E^2\sqrt{-6a\Psi}}  \Bigg)   \Bigg|^{\Psi_B}_{\Psi_A}\,, 
\end{eqnarray}
where we have used that $\Psi \equiv \Psi_2= \frac{1-4a}{24ar^2}$ is the only non-zero  component of the Weyl curvature, in the language of the Newman-Penrose scalars \cite{NPC}, for the Tolman VI 
metric\footnote{Our static and spherically symmetric spacetime metric (\ref{metric})  is of Petrov type D, consequently admitting only one independent non-zero component of the Newman-Penrose Weyl scalars. In our mathematical equations,  we remove the subscript 2 for preventing a cumbersome notation.}. We also recall that this specific Weyl scalar quantifies the strength of gravitational tidal effects of the spacetime it corresponds to \cite{szekeres}.

\begin{figure}
	\begin{center}
		$
		\begin{array}{ccc}
			{\includegraphics[angle=0,scale=0.37]{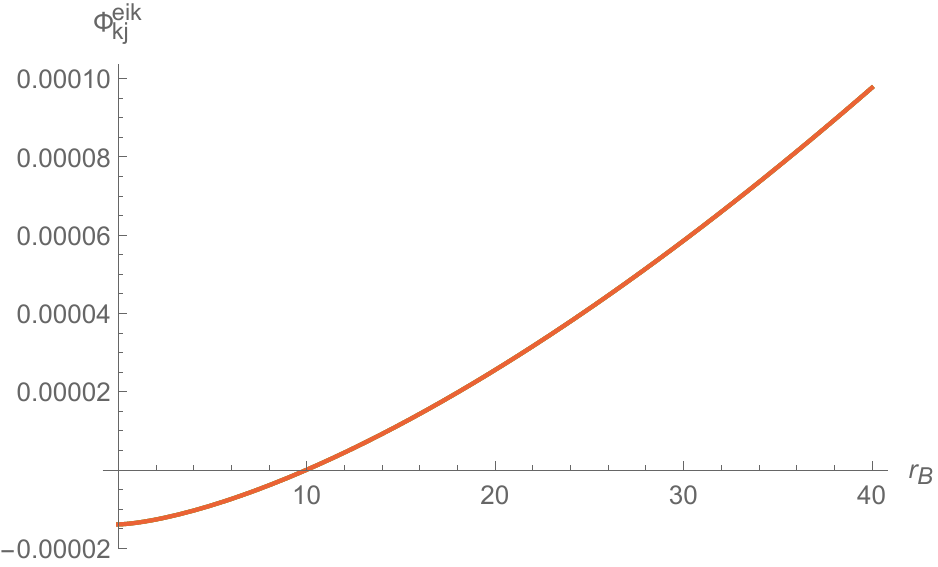}}  &
			{\includegraphics[angle=0,scale=0.37]{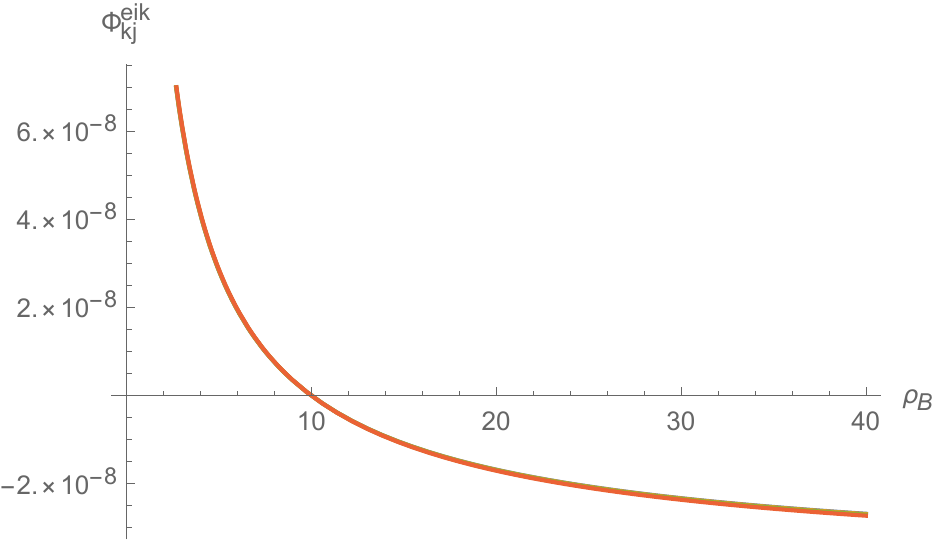}}  &
			{\includegraphics[angle=0,scale=0.37]{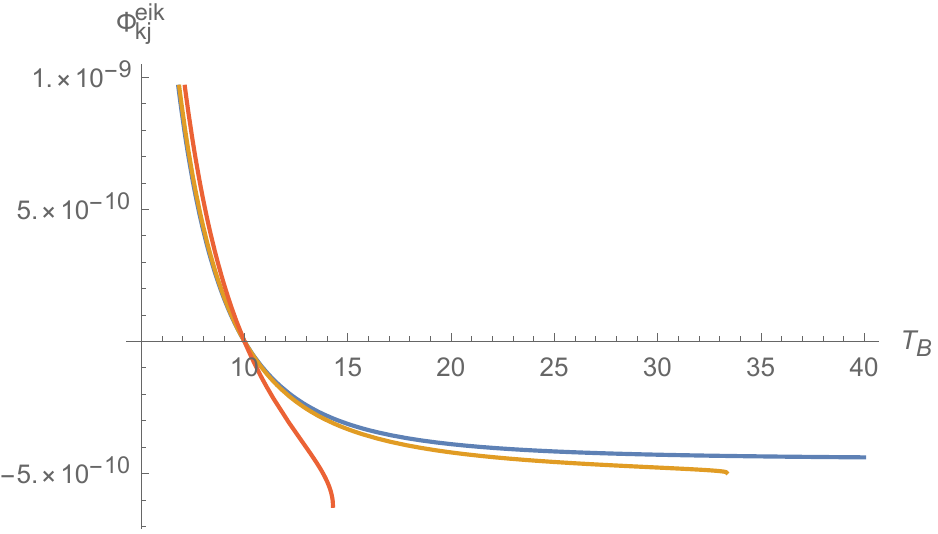}}  \\
			(a) & (b) & (c) 
		\end{array}$
	\end{center}
	\caption{In this figure, we depict the neutrino eikonal phase shift  with respect to the radial coordinate (panel a), background energy density (panel (b)), and temperature of the background matter (panel c) by plotting Eqs.(\ref{finP}), (\ref{eikrho}) and (\ref{eikT}), respectively. The behavior with respect to $\Psi_2$ is expected to be qualitatively similar to that obtained in panel (b) due to the likewise radial decay of $\rho$ and $\Psi_2$. We have chosen $a=\frac{7}{4}$ (implying that our background spacetime is filled by ultrarelativistic matter in gravitational equilibrium), $E=1$, $\Delta m=0.001$, $r_A=\rho_A=T_A=10.0$ (these latter values referring to different spatial regions), $T_\infty=1.0$ and $b=[0.01,\,0.03,\, 0.05,\,0.07]$ for the impact parameter. We obtain, as expected, a piling-up effect with respect to the crossed distance $r_A-r_B$, and decreasing evolutions with respect to $\rho_B$ and $T_B$ with the latter being steeper; quantitative differences with respect to the value of the impact parameter cannot be appreciated by eye, but for its highest value in the context of the temperature study.   }
	\label{fig1}
\end{figure}

In various panels of Fig.\ref{fig1}, we numerically inspect the behavior of the phase shift in the eikonal limit for  neutrinos traveling inside a spacetime supported by ultrarelativistic matter in gravitational equilibrium with respect to {\it (i)} the radial coordinate, {\it (ii)} the background energy density, and {\it (iii)} the  temperature of the background matter. We can identify some converse behaviors between the radial evolution (increasing) and those with respect to the background energy density and temperature, the latter being the steepest. The relationship between eikonal phase shift and Weyl scalar would be similar to that in terms of the background energy density due to the likewise decay $\rho,\, \Psi \propto 1/r^2$. We note that, in general, the considered ranges of independent variables  cover different spatial regions. In this analysis, we find the numerical values of the impact parameter of the geodesic paths to play a minor role in determining the qualitative features of the phase shift.  

\section{Spin effects in the Neutrino oscillations}
\label{ss4}

In the recent paper \cite{frolovA}, the geometric optics treatment for the motion of spinors (such as the Dirac fields) has been refined into a spinoptics formalism: this literature will allow us to work out a second level approximation for the motion of the neutrino wavepackets \cite{4levels}. The basic ingredients of the method are: {\it (i)}  a mass-shell eikonal equation determining the ray trajectory for a high-frequency\footnote{This is consistent with the ultrarelativistic nature of  neutrinos.} spinor wave; {\it (ii)} a transport equation for the spinor polarization; {\it (iii)} a Berry-phase-like correction quantifying the interference between different mass eigenstates. Specifically,  the latter takes into account that when a quantum state is transported around a closed loop in the parameter space, it returns multiplied by a phase determined purely by geometry (in fact, the \lq\lq internal space\rq\rq  --constituted by spin, helicity, polarization, band eigenstate-- of a quantum state evolves slowly as the system moves along a path);  the Berry-like phase, which is geometrically interpreted as the symplectic area enclosed within the curved projective Hilbert space \cite{ANDA},  exhibits the following characteristics \cite{berry1,berry2}: {\it (i)} it does not depend on the time duration of the process, and consequently it is purely geometrical and not dynamical; {\it (ii)} it is gauge-invariant and consequently measurable; {\it (iii)} neutrinos (as any Dirac field) propagating in a curved spacetime acquire geometric Berry-like  phases due to the spinor parallel transport which is set by the spin connection. Thus, the recently developed formalism \cite{frolovA} suits studies of neutrino oscillations because neutrinos: {\it (i)} propagate ultra-relativistically; {\it (ii)} have distinct mass eigenstates; {\it (iii)} acquire relative phases which control flavor oscillations.

In a curved spacetime, the spinoptics phase depends on the geometry through its tetrad $e^a_\mu$ and the corresponding spin connection $\omega^{ab}_\mu=e^a_\nu \nabla_\mu e^{b\nu}$. Since the Tolman VI metric  (\ref{metric}) is diagonal,  it appears natural to choose the following orthonormal tetrad:
\beq
\label{tetrad}
e^{\hat 0}_t=\sqrt{r},\qquad e^{\hat 1}_r=\sqrt{a},\qquad e^{\hat 2}_\theta=r,\qquad e^{\hat 3}_\phi=r \sin \theta\,.
\eeq
The spin connection  built from this tetrad generates the gravitational coupling to be enforced in the Dirac equation. In fact, the WKB expansion  of a Dirac field is written as $\psi_i \approx e^{iS_i/\varepsilon}(u_i+\varepsilon \chi_i)$,  where $S_i$ is the rapidly varying (eikonal) phase  and $u_i$ is the the leading order spinor amplitude. The latter obeys a spin-transport equation analogous to that for the parallel transport of the polarization for photons, but with the additional Dirac structure \cite{frolovA}:
\beq
P^\mu \nabla_\mu u_i =\frac{i}{4}P^\mu \omega^{ab}_\mu \Sigma_{ab} u_i\,, \qquad \Sigma_{ab}=\frac{1}{2} [\gamma_a,\gamma_b]\,.
\eeq
Moreover, $\varepsilon$ is a formal expansion parameter encoding the short-wavelength limit of the Dirac field: {\it (i)} the order $1/\varepsilon^{2}$ amounts to the mass-shell (eikonal) condition; {\it (ii)} the order $1/\varepsilon$ delivers the transport equation for the spinor polarization; {\it (iii)} the order $\varepsilon^0$ gives corrections that include the spinoptics Berry phase. We remark that the spinoptics formalism becomes relevant in astrophysical and cosmological realms because $\varepsilon\sim \frac{\lambda}{R_c} = \frac{1}{E R_c}$ (this $\lambda$ is the neutrino wavelength and $R_c$ is the  curvature of the background spacetime) is very small. We also recall that the regime of applicability of the spinoptics formalism requires  the eikonal phase  to vary rapidly and the amplitude to vary slowly: hence, assessing the magnitude of the spinoptics contribution with respect to that of the  eikonal term we have computed in the previous section is in order.

Quantitatively, the phase of the spinor describing a neutrino receives an extra contribution, beyond that we have computed so far,  which reads as:
\beq
\label{spinintegral}
\Phi_k=\Phi_k^{\rm eik}+\Phi_k^{\rm spin} \,, \qquad \Phi_k^{\rm spin}=\frac{1}{4}\int \omega^{ab}_\mu  S_{(k)ab} \d x^\mu=\frac{1}{4}\int \omega^{ab}_\mu U^\mu  S_{ab} \d \lambda\,,
\eeq
where the latter is a geometric Berry-like phase, which is function of the spin orientation, in which $S_{(k)ab}=\langle u_k| \Sigma_{ab} |u_k \rangle$ is the expectation value of the Dirac spin generator in the mass eigenstate $k$. The physical interpretation of the presented equation is the following: spacetime curvature modifies neutrino oscillations not just via a redshifting of their momentum but also through a spin-connection phase.

Having in hand a velocity four-vector tangent to the geodesic path like that we have already constructed towards the end of Sect.\ref{ss2}, the spinoptics phase shift becomes the following explicit line integral:
\beq
\label{lineint}
\Phi_{kj}^{\rm spin}=\frac{1}{4} \int_{\lambda_A}^{\lambda_B}\omega^{ab}_{\mu}(S_{ab}^{(k)}(x^\mu)-S_{ab}^{(j)}(x^\mu)) U^\mu(\lambda) \d \lambda=\frac{1}{4} \int_{\lambda_A}^{\lambda_B}\left(\omega^{ab}_{t} \frac{\d t}{\d \lambda} + \omega^{ab}_{r} \frac{\d r}{\d \lambda} +\omega^{ab}_{\phi} \frac{\d \phi}{\d \lambda} \right)(S_{ab}^{(k)}-S_{ab}^{(j)})(x(\lambda)) \d \lambda \,.
\eeq
First of all we should note that if the two eigenstates $k$ and $j$ have the same spin orientations (common helicity\footnote{The helicity is the projection of the spin vector onto the momentum vector.}) then $S_{ab}^{(k)}-S_{ab}^{(j)}=0$ and $\Phi_{kj}^{\rm spin}=0$: a non-zero contribution to the neutrino phase shift at the spinoptics level requires a variation in the transported spin expectation values. Interestingly, this is specifically the case when neutrinos travel inside a curved spacetime  as our (\ref{metric}) because helicity is not  conserved in general; in fact, the spinor field obeys a covariant transport equation (the spin expectation value is governed by the spin connection) while the momentum obeys the geodesic equation: since the Levi-Civita connection couples differently to spin and momentum, their directions do not  remain aligned  along a worldline; in summary, helicity is the contraction of two vectors that are transported by different connections, and therefore they \lq\lq twist" differently \cite{helicity}.

Next, we note that for ultrarelativistic neutrinos whose spin is (anti)parallel to the spatial momentum direction $\hat n^k$  in the tetrad frame,  the only non-zero expectation values stem from the spatial rotation generators
\beq
\label{eqss}
S_{\hat i \hat j}^{(l)}=\varepsilon_{\hat i \hat j \hat k}s^{(l) \hat k}\,, \qquad s^{(l) \hat k} \equiv \pm \hat n^{ \hat k}_{(l) }=\pm \frac{P^{ \hat k}_{(l) }}{|P_{(l) }|}
\eeq
with the algebraic sign set by the helicity. In this case, the timelike boost components $S_{\hat 0 \hat i}$
are suppressed by the smallness of $m/E$ and can be dropped at leading order. This level of approximation is applicable to various scenarios of our interest of neutrinos traveling inside  photons' configurations, either at high temperature and density as in the primordial universe and in the cores of stars, or at low temperature as in today Cosmic Microwave Background. This is because \cite{boost}: {\it (i)} $\Sigma^{0i}$, mixing left and right chiral components, is suppressed for ultrarelativistic chiral fermions; {\it (ii)} Standard Model interactions with photon plasma preserve chirality  due to vector-axial  (V-A) structure of the theory; {\it (iii)} photon scattering does not couple significantly to $\Sigma^{0i}$.

Eventually, the integral to be computed in (\ref{lineint}) reduces to the spatial rotation pieces:
\beq
\Phi_{ij}^{\rm rot}\approx \frac{1}{4} \int_{\lambda_A}^{\lambda_B}\omega^{\hat i \hat j}_{\mu}U^\mu \varepsilon_{\hat i  \hat j \hat k}(s^{(i) \hat k}-s^{(j) \hat k}) \d \lambda \,.
\eeq
We will now show that this contribution however vanishes. Explicitly, the non-zero spin-connection 1-form components which enter the worldline integral (\ref{spinintegral}) and which stem from (\ref{tetrad})  for a (planar)   non-radial  motion are:
\begin{eqnarray}
\omega^{\hat 0 \hat 1}_{t}=\frac{1}{2\sqrt{ar}}\,,  \qquad \omega^{\hat 1 \hat 2}_{\theta}=-\frac{1}{\sqrt{a}} \,, \qquad \omega^{\hat 1 \hat 3}_{\phi}=-\frac{\sin \theta}{\sqrt{a}} \,, \qquad \omega^{\hat 2\hat 3}_{\phi}=-\cos \theta \,, \qquad 
 \omega^{ab}_{\mu}=-\omega^{ba}_{\mu}\,.
\end{eqnarray}
Therefore, we obtain
\beq
\Phi_{ij}^{\rm rot}=  \frac{1}{4} \int_{r_A}^{r_B}\left[-\frac{1}{\sqrt{a}} \frac{\d \theta}{\d r} \Delta s^{\hat 3}+\left( \frac{\sin\theta}{\sqrt{a}}\Delta s^{\hat 2}-\cos \theta \Delta s^{\hat 1} \right) \frac{\d \phi}{\d r} \right] \d r= \frac{1}{4\sqrt{a}} \int_{r_A}^{r_B} \frac{\d \phi}{\d r} \Delta s^{\hat 2} \d r\,,
\eeq
where in the last step we have enforced the condition of planar motion on the equatorial plane. Moreover, $\Delta s^{\hat a}(r)=s^{(i)\hat a}(r)-s^{(j)\hat a}(r)$ represents the difference between the local spatial spin expectation values evaluated along the ray at radius\footnote{This computation is carried out in the orthonormal tetrad frame ($\hat 1$, $\hat 2$, $\hat 3$) of mass eigenstate  $i$ and mass eigenstate $j$.} $r$. From (\ref{eqss}) we obtain $\Delta s^{\hat 2}(r) \propto P^\theta =0$. Hence, under our conditions of: {\it (i)} equatorial trajectory, {\it (ii)} helicity eigenstates whose spins are taken to be aligned with the local spatial momentum, {\it (iii)}  leading-order spinoptics approximation, we obtain that the spatial-rotation contribution to the Berry-like phase vanishes exactly. Our result $\Phi_{ij}^{\rm spin}=0$ boils down to that from \cite{heuristic,mash}, in which it has been shown that for ultrarelativistic neutrinos in static spherically symmetric spacetimes, only mass and matter potentials contribute to the oscillations because the SO(3) rotation sector of the spin connection cannot generate any relative phase.

Therefore, we should move to the subleading order of the spinoptics formalism and compute the boost contribution to the phase shift. Specifically, we should reconsider (\ref{spinintegral}), but now picking the boost contributions ($a=0$, $b=\hat i$) \footnote{This $a$ denotes a tetrad index and not the spacetime metric free parameter.}, and change integration variable via $\d \lambda=\d r/K^r$. Eventually  we arrive at
\beq
\label{intL}
\Phi_{ij}^{\rm boost}=  \frac{1}{8a} \int_{r_A}^{r_B} \frac{E}{r\sqrt{E^2 r-L^2-(m r)^2}} \Delta b^{\hat 1}(r) \d r\,,
\eeq
where\footnote{We remark that for a massless Weyl neutrino, the following boost contribution to the phase shift correctly vanishes \cite{Weyloriginal}.}
\beq
\Delta b^{\hat i}(r)=S^{(i)}_{0\hat i}-S^{(j)}_{0 \hat i} \qquad \Rightarrow \qquad  \Delta b^{\hat 1}(r)\approx \Delta b^{\hat 3}(r) \approx \frac{\Delta m_{ij}}{E}  \,,
\eeq
the latter following from
\beq
\langle \Sigma^{\hat 0 \hat i} \rangle \sim \frac{m}{E} \langle \Sigma^{\hat i \hat j}\rangle \sim {\mathcal O}\left( \frac{m}{E}\right)\,.
\eeq
We note that a non-radial motion with $L\neq 0$ would induce a gravitational spin-orbit coupling encoded in the transverse boost component: however, in our configuration, this does not contribute to the phase shift because the geometry (the spin connection) does not \lq\lq push" in that direction since $\omega^{\hat 0 \hat 2}=\omega^{\hat 0 \hat 3}=0$. More precisely, we have
\beq
\label{likewise}
\Delta b^{\hat 1}(r)=\frac{\sqrt{r}\Delta m_{ij}}{E}\cdot \frac{P^{\hat r}}{P^{\hat t}}\,.
\eeq
Therefore, we obtain:
\beq
\label{finboost}
\Phi_{kj}^{\rm spin} \to \Phi_{kj}^{\rm boost}=  \frac{\Delta m_{ij}}{8a^{3/2} E} \int_{r_A}^{r_B} \frac{\d r}{r} = \frac{\Delta m_{ij}}{8a^{3/2} E} \ln \left(\frac{r_B}{r_A} \right)   \,.
\eeq
We should now comment that: {\it (i)} the contribution of the orbital angular momentum $L$ in (\ref{intL}) gets canceled by the likewise factor from (\ref{likewise}); {\it (ii)} the integrand $1/r$ is nothing else than the inverse of the gravitational redshift factor of the spacetime metric (\ref{metric}); {\it (iii)} within both the eikonal and spinoptics phase shifts, whether the neutrinos move inwardly or outwardly does not matter because their direction of motion simply leads to an overall algebraic sign which will get washed away at the level of the conversion probability (as we will discuss in the next section);  {\it (iv)} the integrand entering the eikonal phase shift (\ref{finP0}) exhibits an integrable singularity at the turning point $r=b^2$, while the singularity in $r=0$, corresponding to a curvature blow-up, in the spinoptics phase shift is non-integrable;  {\it (v)} for a radial motion, and at leading order in the ultrarelativistic approximation, we obtain the following condition on the neutrino crossed path for the equality between eikonal and spinoptics phase shifts to hold
\beq
\Phi_{kj}^{\rm eik} \approx \Phi_{kj}^{\rm spin} \qquad \Rightarrow \qquad r_B = \left(\frac{W(\pm \chi)}{4a^2\Delta m_{kj}}\right)^{2/3} \,, \qquad \chi=4a^2 \Delta m_{kj} r_A^{3/2}e^{-4a^2 \Delta m_{kj} r_A^{3/2}}\,,
\eeq
where $W$ denotes the  Lambert special function,  where the algebraic sign depends on whether $r_B>r_A$ or viceversa ({\it e.g.} the solution is unique), and where only the mass difference between two species but not their energy scale is relevant.

Finally, by applying the same mathematical manipulations presented in Sect.\ref{ss3}, this result can be recast in terms of the background energy density, temperature and strength of the pure gravitational field separately as: 
\beq
\label{spinT}
\Phi_{kj}^{\rm spin} \equiv \Phi_{kj}^{\rm boost}=\frac{\Delta m_{ij}}{16a^{3/2} E} \ln \left( \frac{\rho_A}{\rho_B} \right)=\frac{\Delta m_{ij}}{4a^{3/2} E} \ln \left( \frac{T_A}{T_B} \right)=\frac{\Delta m_{ij}}{16a^{3/2} E} \ln \left( \frac{\Psi_A}{\Psi_B} \right)\,.
\eeq
Let us conclude this section by comparing and contrasting the effect of the gravitational field in the eikonal versus spin phase shifts. In general, the contribution to the neutrino phase shift stemming from the geometrical Berry-like phase is suppressed with respect to the eikonal one that we have computed in Sect.\ref{ss3}: in fact,  while they are both dynamical in a curved spacetime setting, they pile up at different rates.  Nevertheless there are situations in which the spin phase shift becomes the leading-order effect; in fact, the spin contribution scales with $\Delta m$ rather than $\Delta m^2$, and it can potentially dominate in regimes where the absolute mass scale is relevant, or in specific gravitational environments  where the spin connection $\omega$ is particularly strong (in our scenario in proximity of the curvature singularity $r=0$ which is not hidden by an horizon). This behavior arises because  gravity does not just affect the path of the neutrinos but  it also twists the internal state of the neutrino flavor in a way that is linearly sensitive to the mass difference. In the next section, we will further confirm these facts by numerically plotting the survival probability of a certain neutrino specie.

\section{Conversion probability}
\label{sectprob}

In this section, we will investigate the conversion probability between two neutrino species exploiting the results for the phase shift that we have obtained in Sects.\ref{ss3}-\ref{ss4}. The two-flavor survival probability hence reduces to:
\beq
\label{probability}
P= 1-\sin^2(2 \theta_{kj}) \sin^2 \left( \frac{\Phi_{kj}}{2}\right) = 1- \sin^2(2 \theta_{kj}) \sin^2 \left( \frac{\Phi_{kj}^{\rm eik}+\Phi_{kj}^{\rm boost}}{2}\right)\,,
\eeq
where this $\theta_{kj}$ is the mixing angle between the two species which can be parameterized as in a flat spacetime \cite{visinelli,referee2}, while we have computed $\Phi_{kj}^{\rm eik}$ in Eq.(\ref{finP}) and $\Phi_{kj}^{\rm boost}$ in Eq.(\ref{finboost}).

Physically, our results allow us to  provide the following different interpretations for the source of the neutrino oscillations: {\it (i)} in the geometrical language, neutrinos oscillate due to a spacetime optical stretching factor $\rho_{NP} \propto 1/r$ as predicted by the Newman-Penrose formalism \cite{NPC}; {\it (ii)} in the gravitational language, the neutrino flavor oscillations stem from the different strengths of the gravitational field at different spatial locations as quantified by the Weyl scalar $\Psi_2$; {\it (iii)} in the matter field language, they are triggered by the spatial gradient of the background energy density  $\rho(r)$; {\it (iv)} in the thermodynamical language,  a  temperature gradient sources the neutrino flavor conversion. We single out the following crucial aspect at the core of each of these physically different interpretations: if $r_A=r_B$,  the values of the previously listed physical quantities do not vary, and  $\Delta \Phi_{kj} =0$ by the very definition of integral.

We should notice that the constant parameter $a$, entering the spacetime metric (\ref{metric}) and fixing the type of background matter content  (\ref{einstein}), acts as an overall proportionality factor for both the eikonal (\ref{finP}) and spin (\ref{finboost}) phase shifts. Its role is converse in the eikonal and spin phase shifts enhancing\footnote{Recall that $a>1$ as from our assessment, based on energy condition and causality requirements, below Eq.(\ref{einstein}).} one and shrinking the other. Thus, its  significance in the context of the neutrino flavor survival probability is to shift the  locations of the maxima/minima of such a probability which  occur at $n \pi$, with $n$ an arbitrary integer number, for the argument of the trigonometric functions in (\ref{probability}). We also notice that the square roots of $a$ are well-posed because $a>0$ as from the Lorentzian signature of the spacetime metric. Furthermore, the conversion probability, --based on either the eikonal or spinoptics phase shift or their combined contributions-- is not sensitive to whether the neutrinos move outwardly or inwardly\footnote{Here we use that $\Phi_{kj}\to -\Phi_{kj}$ as from the antisymmetry of the integral, and that we take the square of the sine which is an odd function of its argument. } (that is climbing up or rolling down the gravitational potential).

In the scenario of low temperature\footnote{Note that this indeed means far away from the curvature singularity in whose proximity instead spin effects can become noticeable as identified at the end of the previous section.} and at the leading order of ultrarelativistic neutrinos, Eq.(\ref{eikT}) reduces to that for an empty flat spacetime background \cite{heuristic}
\beq
\Phi_{kj}^{\rm eik} \to \frac{\sqrt{a}\Delta m^2}{3E}\left(\frac{T_\infty}{T} \right)^3 \bigg|_{T_A}^{T_B}=\frac{\sqrt{a}\Delta m^2}{3E}r^{3/2} \bigg|_{r_A}^{r_B}=\frac{ \Delta m^2 R_{\rm eff} }{2E}\,,
\eeq
in which the role of the orbit's impact parameter $b$ is washed away. Here, we have used Tolman's law for the redshift of the temperature with respect to the strength of the gravitational field and Eq.(\ref{Reff}) providing the effective distance length in our curved spacetime. Consequently, neutrinos traveling inside today Cosmic Microwave Background Radiation oscillate as if they were moving inside an empty flat spacetime; our result is consistent with the analysis from \cite{low} in which, jointly utilizing {\it (i)} either real-time or imaginary-time finite-temperature field theory, {\it (ii)} Standard Model weak interactions theory, and {\it (iii)} one-loop neutrino self-energy diagrams, it was found that neutrino oscillations are sensitive only to flavor-dependent refractive effects\footnote{\lq\lq Thermal refractive effects"  refer to the modification of the phase velocity and dispersion relation experienced by neutrinos caused by coherent forward scattering of particles in a thermal medium, even when no actual collisions occur.}: indeed low-temperature photon backgrounds do not generate them. Moreover, restricting ourselves to the eikonal contribution to the phase shift,  the survival probability is maximal/minimal  for path lengths, in terms of the radial coordinate, such that $r_B^{3/2}-r_A^{3/2}=\frac{3n\pi E}{\sqrt{a}\Delta m^2}$ and $r_B^{3/2}-r_A^{3/2}=\frac{3n\pi E}{2\sqrt{a}\Delta m^2}$, respectively.

Our heuristic treatment captures also the correct asymptotic behavior of the neutrino flavor survival probability at high temperature regimes. Recalling the negligible role played by the orbit's impact parameter as clarified in Fig.\ref{fig1}, and consequently considering a radial motion in Eq.(\ref{eikT}), we have $\lim_{T \to \infty} \Phi_{kj}=0$ implying that the eikonal survival probability tends to one.  This result recovers that from thermal quantum field arguments \cite{dolgov,high}: in the language of the latter, the effective dispersion relation of the neutrinos would acquire a  thermal self-energy term suppressing the oscillation frequency which consequently becomes very slow; the quantum-mechanical scrambling, which dominates at high temperature, and the frequent scattering of neutrinos  generate a quantum damping rather than a flavor conversion. It should be appreciated that our procedure prevents the computational difficulties arising in the quantum field theory approach: specifically, at high temperature,  neutrinos cannot be treated as isolated particles, because they become quasiparticle excitations of a thermal plasma requiring the computation of thermal propagators rather than wavefunctions \cite{low}; moreover, resuming these various contributions is difficult because perturbative schemes fail at high density \cite{diff}. However, the inclusion of spinoptics effects (\ref{spinT}) in the phase shift (\ref{probability}) changes drastically this picture: in the presence of a divergent spin-boost phase $\Phi^{\rm boost} \to \infty$, the flavor-frozen $P \to 1$ limit is replaced by the Gribov-Pontecorvo  regime of rapid phase averaging for which we should consider\footnote{This rapid phase accumulation effectively \lq\lq integrates" the $\sin^2 \Phi$ term over many periods within an infinitesimally small radial interval \cite{OHL}.} $\sin^2 \Phi \to \langle\sin^2 \Phi \rangle =\frac{1}{2}$ \cite{pontecorvo2}. Therefore, at the naked singularity of the Tolman VI metric, the gravitational field becomes so strong that  all flavor information is lost, resulting in an equal probability of measuring any flavor.

\begin{figure}
	\begin{center}
		$
		\begin{array}{ccc}
			{\includegraphics[angle=0,scale=0.37]{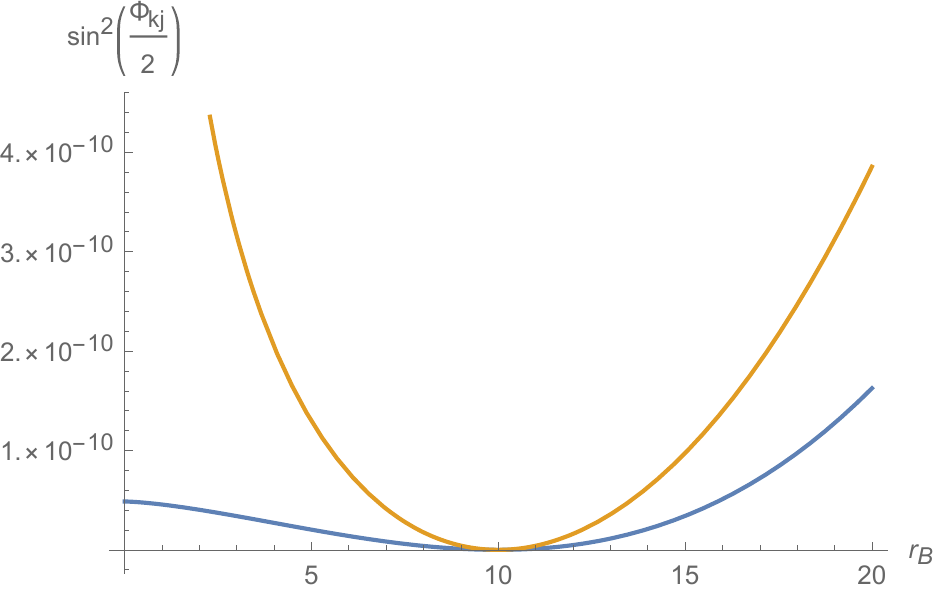}}  &
			{\includegraphics[angle=0,scale=0.37]{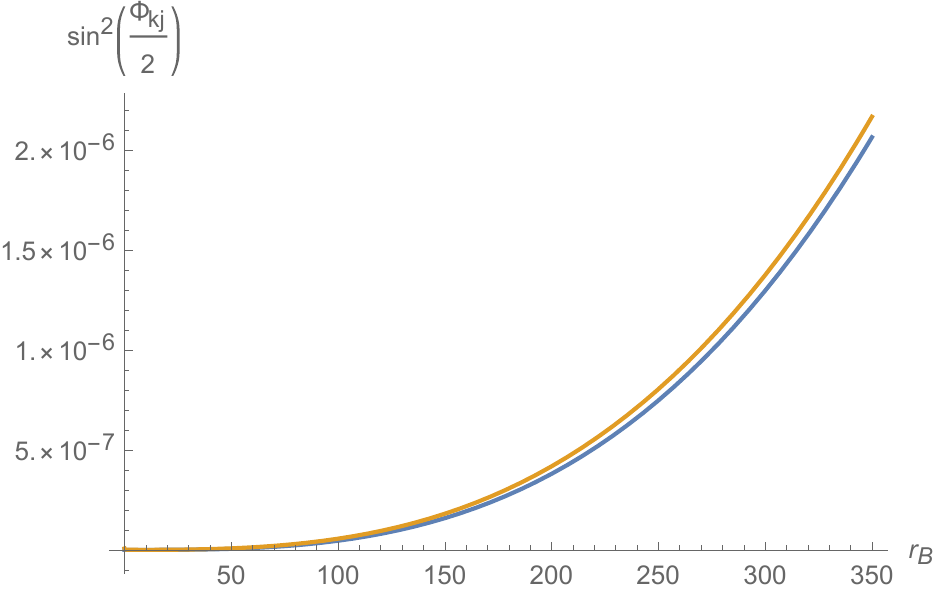}}  &
			{\includegraphics[angle=0,scale=0.37]{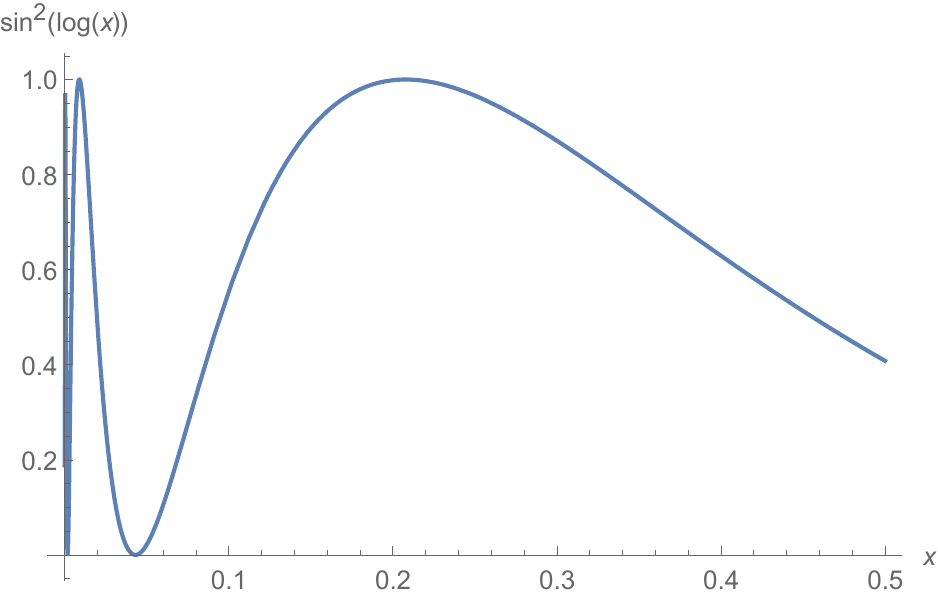}}\\
			(a) & (b)  & (c)
		\end{array}$
	\end{center}
	\caption{In this figure, we depict the radial evolution of the factor $\sin^2 \frac{\Phi_{kj}}{2}$ entering  the neutrino flavor survival probability (\ref{probability}) considering only the eikonal term (\ref{finP}) in the blue curve and the sum of both the eikonal and spin-boost contributions (\ref{finboost}) in orange. We choose the same numerical values for the parameters as in Fig.\ref{fig1} and fix $b=0.05$ (since we have already assessed the role of the orbit's impact parameter). This numerical analysis confirms that the goodness of the eikonal approximation improves --as long as the neutrinos do not hit the central curvature singularity-- as the neutrinos cross longer distances: in fact, the curves, initially separated in panel (a), then get superimposed in panel (b). In panel (c), we instead zoom the region in proximity of the curvature singularity, and we identify the rapidly varying behavior of the spinoptics phase shift which will require the development of an appropriate averaging formalism.  }
	\label{fig2}
\end{figure}

In Fig.\ref{fig2}, we compare and contrast the properties of the position-dependent factor $\sin^2 \frac{\Phi_{ij}}{2}$ entering the neutrino flavor conversion probability at the two different levels of approximation we have worked out in Sects.\ref{ss3}-\ref{ss4}. In fact, we present separately only the eikonal contribution to the phase shift and then the combined eikonal and spin-boost terms: this analysis confirms numerically that indeed the accuracy of the eikonal approximation --as long as the location of the curvature singularity is avoided by the neutrino motion-- improves as the crossed distance increases because its dynamical (rather than geometrical) effects pile up. We also show the rapidly varying behavior of the spinoptics phase shift --at the location of the curvature singularity-- which has called for the averaging procedure discussed above drastically changing the estimate of the neutrino flavor conversion probability in the regime of diverging temperature. 

We should comment that our results for the neutrino oscillation phase and conversion probability we have eventually presented in this section are independent of whether the spacetime is supported by regular or exotic matter. In fact, along the radial direction, by keeping in mind Eq.(\ref{einstein}), we can have $p_r<0$  for $a>2$ but preserving both the correct spacetime causality ({\it e.g.} Lorentzian signature of the  metric) and the energy conditions $\rho>0$ and $\rho+p_r>0$: our results for the neutrino flavor conversion follow solely from the characteristics of the geodesic motion and the geometrical symmetries of the configuration. On the other hand, our findings apply only as long as the coherence of the neutrino wavepackets is maintained. For ultrarelativistic neutrinos, the local group-velocity difference is \cite{coh}
\beq
\Delta v(r) \approx \frac{\Delta m^2}{2E^2_{\rm loc}(r)}=\frac{r \Delta m^2}{2 E^2_\infty}\,.
\eeq
For example, for radial motion, decoherence occurs when the accumulated separation between two wavepackets, after propagating from $r_A$ to $r_B$, becomes of the same order of magnitude of the dispersion in the wavepackets \cite{coh}:
\beq
\sigma_x \sim \Delta R_{\rm eff}(r)=\int_{r_A}^{r_B} \Delta v(r) \d R_{\rm eff}=\sqrt{a}\int_{r_A}^{r_B}  \frac{r \Delta m^2}{2 E^2_\infty}\d r=\frac{\sqrt{a}\Delta m^2 (r^2_B-r^2_A)}{(2E_\infty)^2}\,.
\eeq
Then, the coherence length in terms of the radial coordinate is obtained to be
\beq
r^2_{\rm coh}=r^2_A+\frac{(2E_\infty)^2 \sigma_x}{\sqrt{a} \Delta m^2}\,.
\eeq
Therefore, the numerical value of the parameter $a$, establishing the equation of state of the background material, not only sets the positions of the maxima/minima for the neutrino flavor survival probability, but it also determines the coherence length: a larger value of $a$, corresponding to a steeper decay of the tangential pressure, delivers a shorter coherence length.  We should note also that the velocity difference $\Delta v(r)$ increases/decreases when moving away/towards the curvature singularity which generates a compression effect: the separation between wavepackets for infalling neutrinos  slow down and the gravitational blueshift acts as a coherency preserver. Thus, the neutrinos which hit the curvature singularity are physically overlapped (because $\Delta v(r) \to 0$), but quantum-mechanically  scrambled (because $\Phi^{\rm spin-boost} \to \infty$). Consequently, this feature of the coherence length breaks the symmetry on the neutrino direction of motion which instead characterizes their flavor conversion probability.

\section{Interpretation of the background equation of state, and resonances}
\label{ss5}

The phenomenon of resonance in neutrino oscillations occurs at the point where the effective in-medium mixing angle $\theta_{\rm m}$ makes the flavor conversion extremely efficient \cite{heuristic}. While in a flat spacetime the condition for the Mikheyev-Smirnov-Wolfenstein resonance is $\Delta m^2 \cos(2 \theta_{\rm m})=2EV_{\rm matter}$ \cite{MSW}, in a curved spacetime scenario we should take into account the redshift factors for the local neutrino energy and the background matter potential. In this section, we will examine some physically different and complementary  situations; specifically, for the decay of the background matter potential we will assume separately: {\it (i)} an external test matter field in the gravitational field (\ref{metric})  versus {\it (ii)}  self-gravitating particles (\ref{einstein}). Regarding the type of material in gravitational equilibrium generating the spacetime curvature, we will compare and contrast the following two possibilities of realizing the equation of state $p=\frac{\rho}{3}$ as for relativistic states:  {\it (i)} a photon gas versus  {\it (ii)} ultrarelativistic electrons-positrons with Fermi momentum and temperature much higher than their rest mass $p_F,\, T \gg m_e$.

In a static and spherically symmetric spacetime like (\ref{metric}), the mixing angle in matter reads as \cite{heuristic}
\beq
\tan (2 \theta_{\rm m}(r))=\frac{\Delta m^2 \sin(2\theta_{kj})}{\Delta m^2 \cos(2\theta_{kj})-2E_{\rm loc}(r)V_{\rm loc}(r)}\,,
\eeq
whose determining quantities are related to the Hamiltonian in the frame of a locally static observer as
\beq
\label{hamiltonian}
H=\frac{\Delta m^2}{4E_{\rm loc}(r)}\begin{pmatrix} -\cos(2\theta_{kj})     & \sin(2\theta_{kj}) \\
	\sin(2\theta_{kj})	  &  \cos(2\theta_{kj})
\end{pmatrix} + \begin{pmatrix} V_{\rm loc}(r)    & 0 \\
0	  &  0
\end{pmatrix}\,.
\eeq
The phenomenon of resonance is triggered by the scattering between neutrinos and electrons/positrons: for neutrinos traveling inside a photon gas, the relevant potential $ V_{\rm loc}(r)$ vanishes, and no MSW effect occurs at all, as already established in \cite{low}.

Strong fields such as those in proximity of a curvature singularity generate vacuum polarization phenomena 
\cite{donpage},  while their strong gradient supports a likewise production of neutrinos \cite{strong}. Should we  assume the neutrinos to be traveling through an $e^-e^+$ external test plasma superimposed to the photons supporting the spacetime (\ref{metric}), we should then take into account  the redshift of the local energy of the neutrinos and of the local matter potential with respect to their values measured at spatial infinity \cite{GRSN}:
\beq
\label{local1}
E_{\rm loc}(r)=\frac{E_\infty}{\sqrt{r}}\,, \qquad V_{\rm loc}(r)=\sqrt{r}V_\infty \,.
\eeq
Let us scrutinize these two redshift rescalings.  The former follows from the conservation of energy related to the existence of the Killing Vector Field $\partial_t$ for our considered spacetime metric, while the latter signifies that the superimposed $e^-e^+$ plasma particles pile up towards the center of the configuration. This specific redshift for the local matter potential, possibly generating the resonance, follows from the  Goldreich-Julian density which preserves the electromagnetic quietness of the spacetime (this is the so-called force-free regime) \cite{quiet}: in this formulation, the $e^-e^+$ plasma indeed behaves as a test field with its potential being determined only by the background geometry. The existence of this astrophysical system requires the presence of mutually perpendicular electric and magnetic fields  with the magnetic field admitting  poloidal and toroidal components because  the electric field is purely radial; moreover, the strength of both these fields should be  monotonically decreasing  from the center of the configuration up to spatial infinity. In fact, the density of plasma is set by the divergence of the electric field, which in turn is given by the gradient of the electric potential. Remarkably, the electric potential (\ref{local1}) vanishes in correspondence of the spacetime curvature singularity but the electric field diverges there: this latter behavior  is necessary for  piling up  the plasma via electromagnetic effects in the same spatial region where also the self-gravitating background particles become denser due to their mutual gravitational attraction.

For neutrinos traveling with purely radial velocity, the resonance condition, when considering a superimposed $e^-e^+$ plasma, turns out to be actually independent of the radial position $r$ because it keeps reading the same as that for quantities at spatial infinity in a  flat spacetime\footnote{Keep in mind that our configuration of interest is indeed asymptotically flat as it can be understood from the decays of energy density, pressure, and Newman-Penrose Weyl scalar.}:
\beq
\label{resonanceflat}
\Delta m^2 \cos(2\theta_{kj}(r))=2E_{\rm loc}(r)V_{\rm loc}(r)=2 E_\infty V_\infty\,.
\eeq
This result can be obtained also along the following route. Due to the combined scalings of local energy and local potential (\ref{local1}), the Hamiltonian (\ref{hamiltonian}) becomes $H=\sqrt{r}H_\infty$. Therefore, while the overall energy scale of the neutrino oscillations changes as the neutrino moves (with the oscillations getting \lq\lq faster" or \lq\lq slower"), the mixing angle remains constant throughout the entire trajectory. In other words, the neutrinos remain in the same mixing state in which they were born all along their traveled path, because the geometrical redshift influences the background and the superimposed matter terms identically.

Let us now scrutinize the role of the neutrino's angular momentum in the adiabaticity parameter. We recall that a larger value of the impact parameter means that the trajectory is more tangential, and this circumstance makes the phase accumulate more quickly. Moreover, the adiabaticity parameter $\gamma$ is a quantitative measure of how much slowly the background propagation environment varies (keep in mind that in the Hamiltonian  of the process (\ref{hamiltonian}), $V_{\rm loc}(r)$ is position-dependent) compared with the intrinsic oscillation timescale: the value of the adiabaticity parameter informs us whether a neutrino which begins its motion in some instantaneous propagation eigenstate (matter eigenstate) will remain in that same eigenstate as it moves through space. Quantitatively, the adiabaticity parameter is given by the ratio of the  level splitting with respect to the rate of change of the eigenstates (or, in other words, it is  the ratio of the internal oscillation timescale with respect to the external variation timescale), and, in practical terms, it controls whether the resonant flavor conversion is efficient or suppressed. A high value of the adiabaticity parameter means that the Hamiltonian changes slowly with the neutrino tracking the instantaneous eigenstate and consequently the conversion is favored and noticeable; low adiabaticity means instead that the Hamiltonian changes rapidly with the neutrino jumping between eigenstates and consequently the conversion is suppressed or probabilistic. 
Finally, the probability of a non-adiabatic jump is quantified by the Landau-Zener formula \cite{landauA}:
\beq
\label{zener}
P_{\rm jump}=e^{-\pi \gamma_{\rm res}/2}\,,
\eeq
where at MSW resonance\footnote{The next equation essentially states that
\beq
\gamma_{\rm res}=\frac{({\rm Energy\,\, Gap})^2}{|{\rm Rate\,\, of\,\, Change \,\, of \,\,the\,\, Diagonal \,\,Difference}|} \,,
\eeq
where the denominator refers to the diagonal terms in the Hamiltonian (\ref{hamiltonian}). 	}
\beq
\label{gres}
\gamma_{\rm res}=\left[\left(\frac{\Delta m^2 \sin(2\theta_{kj})}{2E_{\rm loc}}\right)^2 \bigg|\left(\frac{ \operatorname{d} }{\operatorname{d} s}\left( \frac{\Delta m^2 \cos(2\theta_{kj})}{2E_{\rm loc}} -V_{\rm loc}\right) \right)^{-1}\bigg| \right]\Bigg |_{\rm res}\,,
\eeq
in which $s$ denotes the proper (to be contrasted against  coordinate) distance along the neutrino path. The Landau-Zener formula holds for Hamiltonians  which depend on a single monotonically varying parameter, in this case the neutrino position $r$; this Hamiltonian delivers a Schr\"odinger-like transport equation following from the WKB approximation of the Dirac equation. We will now see that a non-radial motion enlarges the proper path length across the resonance region, increasing $\gamma$ and making the flavor conversion more adiabatic than in the purely radial propagation. If the process is slow, the neutrinos remain in their instantaneous eigenstates and the evolution is adiabatic; if it is fast, the system jumps to the other eigenstate and the evolution is non-adiabatic. For the scenario of superimposed $e^-e^+$ plasma previously introduced, we obtain that the transition is always completely adiabatic, regardless of how close the neutrinos approach the central naked singularity, by the very definition of resonance (\ref{resonanceflat}) entering the denominator of the adiabaticity parameter.

For a self-gravitating $e^-e^+$ plasma and redshifted neutrino energy, we should write\footnote{Although this specific physical interpretation requires $a=7/4$, we treat the problem in full mathematical generality without fixing the value of the latter parameter.}
\beq
\label{profile}
V_{\rm loc}(r)=\frac{C}{r^2}\,, \qquad C=\frac{\sqrt{2}G_F Y_e}{m_N}\left(1 -\frac{1}{a}\right)\,, \qquad E_{\rm loc}(r)=\frac{E_\infty}{\sqrt{r}}\,,
\eeq
where $G_F$ is the Fermi constant (the low-energy effective coupling of the weak interaction), $m_N$ is the nucleon mass, and $Y_e$ the electron fraction of the medium. Then, we compute (\ref{gres}) using the chain rule via
\beq
\frac{\d s}{\d r}=\sqrt{a+\left(r \frac{\d \phi}{\d r}\right)^2 }=\sqrt{\frac{ar}{r-b^2}}\,,
\eeq
obtaining 
\beq
\label{finalgamma}
\gamma_{\rm res}=\frac{\sqrt{a}}{10 C}\left(\frac{r^2_{\rm res} \Delta m^2 \sin(2\theta_{kj})}{E_\infty} \right)^2\sqrt{\frac{r_{\rm res}}{r_{\rm res}-b^2}}
\eeq
in which
\beq
r_{\rm res}=\left(\frac{2CE_\infty}{\Delta m^2 \cos(2\theta_{kj})}\right)^{2/5}
\eeq
is the resonance radius. This mathematical formula means that along the neutrino worldline, there exists an event where the  value of the radial coordinate is $r_{\rm res}$ and at that event the local MSW resonance condition is satisfied; we remark that this is a gauge-invariant (and not coordinate-dependent) result as defined from the area of the surface at which it occurs  \cite[Eq.(6.1.3)]{wald}. Via Tolman's law we can easily compute the resonance temperature. Therefore, the neutrino propagation in the Tolman VI metric is characterized by a  resonance location induced by genuine gravitational effects, in contrast to the flat-spacetime case. Let us now comment on the underlying physics: {\it (i)} the density profile (\ref{profile}) relates to the spacetime curvature; {\it (ii)} the angular momentum ({\it e.g.}  the impact parameter $b$), which makes grazing trajectories linger near the resonance via a lensing effect which stretches the resonance region, enhances the Landau-Zener probability of a non-adiabatic jump; {\it (iii)}  spin-rotation effects vanish at the  level of the flavor conversion probability, but matter-induced resonance survives and it is enhanced by geometrical effects. We remark that points  {\it (ii)} and  {\it (iii)} deal with the \lq\lq kinematics of te process" and the \lq\lq symmetry of the interactions", respectively.

We have commented that the $e^-e^+$ plasma interpretation of the material filling our assumed spacetime metric may hold at most at high temperature. As a consequence of Tolman's law, $r_{\rm res}$ should be \lq\lq small" (implying that the resonance is  favored towards the center of the configuration), and therefore $\gamma_{\rm res}$ tends to zero. In this spatial region, the Landau-Zener probability (\ref{zener}) is enhanced; we should now explain that this latter behavior does not automatically follow from the condition of high temperature, but from the sharp gradient of the potential ({\it e.g.} a non-flat Hamiltonian) \cite{landauA}: this is indeed the case when our (\ref{profile})   approaches the center of the system\footnote{These considerations do  not rely on the WKB approximation for the solution of the Dirac equation. Thus, the applicability limitation of a slowly varying gravitational field (with respect to the evolution of the neutrino eigenstates) does not trouble our present analysis. Therefore, this latter assessment is relevant also for spatial regions arbitrarily close to the center $r=0$ and at asymptotically high temperature and density.}. The correspondence between high temperature regime and sharp gradient of the potential is a specific property characterizing the Tolman VI metric (\ref{metric}).

Let us now summarize our different findings about possible resonances for the neutrino oscillations inside a spacetime described by the Tolman VI metric in general relativity. In this analysis, the crucial point is not anymore the geometrical symmetry of the configuration, as for the flavor conversion probability, but the physical interpretation of the type  of particles supporting the equation of state inferred via the Einstein equations. {\it (i)} If we regard $p=\frac{\rho}{3}$ as describing a photon gas, then there cannot be any MSW resonance at all because of the vanishing of the MSW potential; {\it (ii)} there would instead be a MSW resonance in an $e^-e^+$ plasma composed by particles whose kinetic energy dominates over their rest mass for a Fermi momentum  such that $p_F \sim T \gg m_e$. Our same result  {\it (i)} was obtained for low and high temperature --separately characterizing the present-day and early-time Cosmic Microwave Background-- in \cite{low} and \cite{dolgov}, respectively. Scenario {\it (ii)} can deal with the description of  the primordial universe (for which interpretation we refer to \cite{dolgov} for a review of previous studies of MSW resonances) or neutron star crusts and outer cores at high density \cite{bethe}, locally consistent with the spacetime metric we have assumed. We recall that the proposal of the Tolman VI metric  for modeling stars dates back to the seminal paper \cite{tov1}, while for its previous adoptions in the description of a radiation-dominated universe we refer to \cite[p.2674]{miotolman} and \cite[p.733]{MTW}. We should  remark also that the treatment of case {\it (i)} is independent of both the specific details of the gravitational effects encapsulated in our spacetime metric and of our assessment of general relativistic  effects on the neutrino propagation; for the quantitative  results about {\it (ii)}, we should instead explicitly adopt the redshift factor predicted by the spacetime (\ref{metric}) via the Einstein field equations: this constitutes our own original contribution for this section.

\section{Conclusions}
\label{ss6}

In this paper, we have studied the phenomenon of neutrino oscillations when traveling inside a spacetime described by the Tolman VI metric (\ref{metric}). The latter is a one-parameter family of solutions of the Einstein field equations in which, for a specific choice of the free parameter, the pressure becomes isotropic and the stress-energy tensor simultaneously tracefree. For this subcase, we can interpret the spacetime to be supported by a gas of massless bosons such as photons self-gravitating without preferred directions, or ultrarelativistic fermionic plasma $e^-e^+$ with global charge neutrality. The relevance of the equation of state $p=\frac{\rho}{3}$ for the description of fermions at high pressure, for example at the centers of stars, was already suggested in Tolman's seminal paper \cite{tov1}. Last but not least, this set-up of a ultrarelativistic equation of state can suit also the modeling of neutrino oscillations inside their own generated fields. In this paper, we have provided exact analytical closed-form results for the phase oscillations of neutrinos, also allowing for an arbitrary value of the impact parameter characterizing the neutrino orbits. We have performed a comparative analysis  between the predictions of the eikonal limit in the geometric optics approximation of the neutrino wavepackets against the roles of the spin effects by taking advantage of the  recently developed spinoptics formalism \cite{frolovA}; regarding this latter level of description, we have discussed that purely geometrical considerations, on our static and spherically symmetric configuration, rules out possible  contributions to the neutrino phase oscillation from spatial rotation generators  permitting only those from boosts. Then, we have numerically confirmed that, apart inside a small neighborhood of the naked singularity,  the spinoptics effect in the flavor conversion probability is  suppressed with respect to the eikonal  term. Moreover,  we have certified the correctness of our results by explicitly checking their asymptotic limits at low and high temperature and comparing with literature based on quantum field theory rather than heuristic arguments;  inside a low temperature photon gas the neutrinos oscillate as in an empty flat spacetime, while at high temperature their eikonal\footnote{We prefer to single out some paragraphs below the careful inspection of the precise signature of the  location of the naked singularity itself in the spinoptics approximation of the flavor survival probability.} flavor survival probability tends to one. It should nevertheless  be appreciated that our results are more general than  those previous literature thanks to  the presence of an arbitrary spacetime metric parameter $a$ which accounts for different types of material filling the spatial region inside which the neutrinos travel. Lastly, we have discussed that the physical interpretation of the nature of the particles realizing the equation of state $p=\frac{\rho}{3}$ becomes crucial for establishing whether resonances can occur; in fact, inside a photon gas they are forbidden, while we have compared and contrasted the two cases of electron-positron plasma, either superimposed as a test field or self-gravitating, by identifying the roles of the redshift factor and of the decay of the potential. In fact, while neutrino oscillations could be triggered by gravitational interactions encoded in the Christoffel symbols of the neutrino geodesic motion, resonance phenomena are a consequence of electroweak interactions.

On the other way round, we have not only corroborated the applicability of the heuristic treatment of neutrino phenomena via a specific quantitative analysis, but we have also elucidated some properties of the Tolman VI metric. First of all, we have clarified that a specific value of the spacetime metric free parameter makes  the stress-energy tensor simultaneously conformal invariant (because tracefree) and  invariant under spatial rotations in all directions: for $a=7/4$ in  (\ref{metric}), which admits the timelike Killing Vector Field $\partial_t$, the symmetry group is $SO(3,1) \ltimes {\rm Conf}$, while for generic values of $a$  the Lorentz symmetry is broken by the anisotropy of the matter content \cite{miotolman}. In fact, this  model free parameter, determining the equation of state of the material with which the neutrinos interact, also sets both the positions of the maxima/minima of the flavor conversion probability and the coherence length; however, its value generating maximal symmetry at the levels of spatial geometry and matter content does not appear to bear specific signatures in  the flavor conversion probability. Specifically, we have explained that the case of exotic matter with negative radial pressure does not trouble our general considerations for the phase oscillation because they follow entirely from the features of the geodesic motion and from the geometrical symmetries of the configuration. On the other hand, the meaning carried by the equation of state of the material filling the space becomes relevant for the correct and consistent prediction of MSW resonances. We should also emphasize  that our results do not constitute coordinate artifacts because they are genuine gauge-independent outcomes as noticed by re-interpreting the radial coordinate as related to the area of the surface on which the phenomena take place \cite[Eq.(6.1.3)]{wald}.   Therefore, our present work belongs to the class of research clarifying the separated  roles of \lq\lq geometrical" versus \lq\lq physical" properties of the background environment in triggering the phenomenon of neutrino oscillations and resonances; those can therefore be interpreted as kinematical and dynamical processes, respectively: specifically, the equation of state of the supporting matter field can be derived from the spacetime metric tensor only choosing a gravitational paradigm.  In other words, from the theoretical perspective,  we have investigated the consequences of the presence of pure gravitational versus pure matter fields; unlike in Newtonian physics, in the realm of relativistic theories of gravitation, the latter do not anymore coincide with each other, as already established in the context of black hole entropy \cite{entropy}.

In Sect.\ref{sectprob}, we have explicitly checked that our heuristic formalism recovers the qualitative behaviors predicted by a quantum field theory treatment of the conversion probability in the asymptotic regimes of both low and high density and temperature. In the former case the neutrinos oscillate as if they were traveling inside an empty flat spacetime, while in the latter the eikonal survival probability tends to one;  it should be emphasized that our procedure avoids complicated resummations customarily arising in thermal quantum field theory because it completely circumvents the adoption of perturbative schemes. These two limiting analyses cover the scenarios of neutrino oscillations in today and early-time Cosmic Microwave Background and inside star cores which had already been studied in the literature we have compared with. On the other hand, our analysis  has also revealed  the importance of accounting  for the spinoptics phase shift, which scales linearly with the mass difference between two species, when neutrinos arrive at a curvature singularity (which in the Tolman VI metric  corresponds to a region of infinite temperature): the flavor conversion probability is not anymore locked but there is an equal probability of finding the neutrinos in any possible eigenflavor.  

Our present research complements previous literature trying to observationally discriminate naked singularities from black holes by either looking at the images of their accretion disks \cite{naked1}, echoes of perturbations \cite{naked2}, and radiative signatures \cite{naked3}. In fact, in Sect.\ref{sectprob}, we have derived the following precise prediction relying on neutrino Physics for testing the actual existence in our Universe of naked singularities described according to the Tolman VI metric: in that curvature regime the neutrinos are simultaneously in flavor-equilibrium, physically overlapped, and quantum-mechanically scrambled; the wavepackets are perfectly overlapped, allowing them to interfere, but the (boost spinoptics) phase is varying so rapidly that any physical detector  can see only the mean value of the oscillations. We emphasize that the latter does not constitute a resonance phenomenon because the scrambling is a consequence of the geometrical Thomas precession \cite{Thomas}: the divergent gravitational acceleration the neutrinos experience in proximity of the curvature singularity averages the flavor conversion probability towards $1/2$.

Further specific quantitative predictions for the phase shift can arise at intermediate energy scales as in the cosmic epoch after  annihilation and in the envelope of supernovae. Let us conclude focusing on the latter. Neutrinos can be trapped inside the very dense regions of neutron stars \cite{stul} delaying their cooling down according to the Urca and bremsstrahlung processes \cite{urca}: neutrinos in fact influence the thermal evolutions of stars by possibly carrying away some amount of energy \cite{zeldovich}. In future projects, we intend to perform the astrophysical analysis of such phenomenon relying on our established theoretical formalism of coherent neutrino propagation via kinematical equations. For the time being,  we note that our obtained equal probability is consistent with the result from state-of-the-art simulations of the dynamics of neutrino's slow modes \cite{SNA}. Moreover, we provide the following order-of-magnitude estimate.  The envelope of a supernova can be treated as a scale-invariant radiation bath consistent with Stefan-Boltzmann law in a region of $R \sim {\mathcal O}(2)$ km at a temperature of $T\sim {\mathcal O}(10)$ K, and we should assume $E={\mathcal O}(1)$ MeV and $\Delta m^2 ={\mathcal O}(-4)$ eV$^2$ \cite{supernovarep}; for a spherical collapse such that $b=0$, we can provide a preliminary result of\footnote{Here we use that 1 eV= ${\mathcal O}(-19)$ J, $G={\mathcal O}(-10)$ m$^3$kg$^{-1}$s$^{-2}$, $c={\mathcal O}(8)$ m/s, $\hbar={\mathcal O}(-34)$ m$^2$s$^{-1}$kg. We also note that we have restored International System units so that the phase becomes dimensionless, and that energy conditions on the background material restrict $1<a<2$.} $\Phi^{\rm eik}_{kj} \sim (cG\hbar)^{-1/4} \frac{\sqrt{a} \Delta m^2 R^{3/2}}{3 \hbar E}\sim \sqrt{a}{\mathcal O}(22)$:  as a consequence of combined effects from gravity $G$, special relativity $c$ and quantum mechanics $\hbar$, the neutrinos propagate in an averaged oscillation regime, meaning that the flavor conversion is governed by the mixing angles rather than by the actual distance traveled.

\section*{Acknowledgments}
DG is a member of the GNFM working group of the Italian INDAM.

\section*{Data and code availability statement}
This is a purely theoretical work with no new data, and not involving any simulation either.

\section*{Conflict of interest declaration}
The author declares that he has no known competing financial interests or personal relationships that could have appeared to influence the work reported in this paper.


\begin{thebibliography}{200} 
	
	
\bibitem{dolgov}
Alexander D. Dolgov, ``Neutrinos in cosmology'',
{\hypersetup{urlcolor=vividviolet}\href{https://www.sciencedirect.com/science/article/abs/pii/S0370157302001394?via%3Dihub}{Phys. Rept.  \textbf{370} (2002)  333}}, \href{https://arxiv.org/abs/hep-ph/0202122}{[arXiv:hep-ph/0202122]}.	
	
\bibitem{stars}
Alessandro Mirizzi, Irene Tamborra, Hans-Thomas Janka, Ninetta Saviano, Kate Scholberg, Robert Bollig, Lorenz Hudepohl, Sovan Chakraborty, ``Supernova Neutrinos: Production, Oscillations and Detection", {\hypersetup{urlcolor=vividviolet}\href{https://www.sif.it/riviste/sif/ncr/econtents/2016/039/01-02/article/0}{Riv. Nuovo Cim.  \textbf{39} (2016)   1}}, \href{https://arxiv.org/abs/1508.00785}{[arXiv:astro-ph.HE/1508.00785]}; Lucas Johns, Sherwood Richers, Meng-Ru Wu, ``Neutrino Oscillations in Core-Collapse Supernovae and Neutron Star Mergers", {\hypersetup{urlcolor=vividviolet}\href{https://www.annualreviews.org/content/journals/10.1146/annurev-nucl-121423-100853}{Annu. Rev. Nucl. Part. Sci.   \textbf{75} (2025)   399}}, \href{https://arxiv.org/abs/2503.05959}{[arXiv:astro-ph.HE/2503.05959]}; Maria Cristina Volpe, ``Neutrinos from dense environments: Flavor mechanisms, theoretical
approaches, observations, and new directions", {\hypersetup{urlcolor=vividviolet}\href{https://journals.aps.org/rmp/abstract/10.1103/RevModPhys.96.025004}{Rev. Mod. Phys.   \textbf{96} (2024)  025004}}, \href{https://arxiv.org/abs/2301.11814}{[arXiv:hep-ph/2301.11814]}.

\bibitem{carlos}
Remo Ruffini, Carlos Ra\'ul Arg\"uelles, Jorge Armando Rueda, ``On the core-halo distribution of dark matter in galaxies", {\hypersetup{urlcolor=vividviolet}\href{https://academic.oup.com/mnras/article/451/1/622/1345772?login=false}{Mon. Not. Roy.  Astron. Soc.   \textbf{451} (2015)   622}}, \href{https://arxiv.org/abs/1409.7365}{[arXiv:astro-ph.GA/1409.7365]}.

\bibitem{infDM}
Mirzabek Alloqulov, Ahmadjon Abdujabbarov, Bobomurat Ahmedov, Chengxun Yuan, ``Can a Dehnen-type dark matter halo affect the neutrino flavor oscillations?", {\hypersetup{urlcolor=vividviolet}\href{https://link.springer.com/article/10.1140/epjc/s10052-026-15442-4}{Eur. Phys. Jour. C   \textbf{86} (2026)   208}}, \href{https://arxiv.org/abs/2510.20563}{[arXiv:gr-qc/2510.20563]}.

\bibitem{pontecorvo1}
Bruno Pontecorvo, ``Inverse Beta Processes and Nonconservation of Lepton Charge",   {\hypersetup{urlcolor=vividviolet}\href{https://inspirehep.net/literature/42736}{Sov. Phys. JETP \textbf{7} (1958) 172; Zh. Eksp. Teor. Fiz. \textbf{34} (1957) 247}}.

\bibitem{pontecorvo2}
Vladimir Gribov, Bruno Pontecorvo, ``Neutrino astronomy and lepton charge",   {\hypersetup{urlcolor=vividviolet}\href{https://www.sciencedirect.com/science/article/abs/pii/0370269369905255?via%3Dihub}{Phys. Lett. B \textbf{28} (1969) 493}}.
	
\bibitem{gellmann}
Murray Gell-Mann, Abraham Pais, ``Behavior of Neutral Particles under Charge Conjugation",   {\hypersetup{urlcolor=vividviolet}\href{https://journals.aps.org/pr/abstract/10.1103/PhysRev.97.1387}{Phys. Rev. \textbf{97} (1955) 1387}}.
	
\bibitem{majorana}
Ettore Majorana, ``Teoria simmetrica dell'elettrone e del positrone",   {\hypersetup{urlcolor=vividviolet}\href{https://link.springer.com/article/10.1007/BF02961314}{Il Nuovo Cimento \textbf{14} (1937) 171}}.	

\bibitem{benattiA}
Fabio Benatti, Roberto Floreanini, ``Massless neutrino oscillations", {\hypersetup{urlcolor=vividviolet}\href{https://journals.aps.org/prd/abstract/10.1103/PhysRevD.64.085015}{Phys. Rev. D   \textbf{64} (2001)  085015}}, \href{https://arxiv.org/abs/hep-ph/0105303}{[arXiv:hep-ph/0105303]}.	
	
	
	
\bibitem{exp1}
The {\it Super-Kamiokande Collaboration}, Yasutaka Fukuda {\it et al.}, ``Evidence for oscillation of atmospheric neutrinos", {\hypersetup{urlcolor=vividviolet}\href{https://journals.aps.org/prl/abstract/10.1103/PhysRevLett.81.1562}{Phys. Rev. Lett.   \textbf{81} (1998)   1562}}, \href{https://arxiv.org/abs/hep-ex/9807003}{[arXiv:hep-ex/9807003]}.	

\bibitem{exp2}
The {\it SNO Collaboration}, ``Measurement of the rate of $\nu_e + d \to p + p + e^-$ interactions produced by $^8 B$ solar neutrinos at the Sudbury Neutrino Observatory", {\hypersetup{urlcolor=vividviolet}\href{https://journals.aps.org/prl/abstract/10.1103/PhysRevLett.87.071301}{Phys. Rev. Lett.  \textbf{87} (2001)   071301}}, \href{https://arxiv.org/abs/nucl-ex/0106015}{[arXiv:nucl-ex/0106015]}.


\bibitem{bilenky}
Samoil M. Bilenky, Carlo Giunti, J.A. Grifols, Eduard Mass\'o, ``Absolute Values of Neutrino Masses: Status and Prospects", {\hypersetup{urlcolor=vividviolet}\href{https://www.sciencedirect.com/science/article/abs/pii/S0370157303001029?via%3Dihub}{Phys. Rept.   \textbf{379} (2003)  69}}, \href{https://arxiv.org/abs/hep-ph/0211462}{[arXiv:hep-ph/0211462]}.
		
\bibitem{american}
Palash B. Pal, ``Dirac, Majorana and Weyl fermions", {\hypersetup{urlcolor=vividviolet}\href{https://pubs.aip.org/aapt/ajp/article-abstract/79/5/485/1057660/Dirac-Majorana-and-Weyl-fermions?redirectedFrom=fulltext}{Am. J. Phys.  \textbf{79} (2011)   485}}, \href{https://arxiv.org/abs/1006.1718}{[arXiv:hep-ph/1006.1718]}.

\bibitem{juno}
Fengpeng An, {\it et al.},  ``Neutrino Physics with JUNO", {\hypersetup{urlcolor=vividviolet}\href{https://iopscience.iop.org/article/10.1088/0954-3899/43/3/030401}{Jour. Phys. G  \textbf{43} (2016)   030401}}, \href{https://arxiv.org/abs/1507.05613v2}{[arXiv:physics.ins-det/1507.05613]}.

\bibitem{heuristic}
Christian Y. Cardall, George M. Fuller, ``Neutrino oscillations in curved spacetime: A heuristic treatment", {\hypersetup{urlcolor=vividviolet}\href{https://journals.aps.org/prd/abstract/10.1103/PhysRevD.55.7960}{Phys. Rev. D  \textbf{55} (1997)   7960}}, \href{https://arxiv.org/abs/hep-ph/9610494}{[arXiv:hep-ph/9610494]}.

\bibitem{slit}
Carlo Giunti, ``The Phase of Neutrino Oscillations", {\hypersetup{urlcolor=vividviolet}\href{https://iopscience.iop.org/article/10.1238/Physica.Regular.067a00029}{Phys. Scripta  \textbf{67} (2003)   29}}, \href{https://arxiv.org/abs/hep-ph/0202063v1}{[arXiv:hep-ph/0202063]}.

\bibitem{visinelli}
Luca Visinelli, ``Neutrino flavor oscillations in a curved space-time", {\hypersetup{urlcolor=vividviolet}\href{https://link.springer.com/article/10.1007/s10714-015-1899-z}{ Gen. Rel. Grav.   \textbf{47} (2015)   62}}, \href{https://arxiv.org/abs/1410.1523}{[arXiv:gr-qc/1410.1523]}.

\bibitem{burgard}
Dharam Vir Ahluwalia, C. Burgard, ``Gravitationally Induced Neutrino-Oscillation Phases", {\hypersetup{urlcolor=vividviolet}\href{https://link.springer.com/article/10.1007/BF03218936}{Gen. Rel. Grav.  \textbf{28} (1996)   1161}}, \href{https://arxiv.org/abs/gr-qc/9603008}{[arXiv:gr-qc/9603008]}.

\bibitem{app}
Leo Stodolsky, ``Matter and light wave interferometry in gravitational fields", {\hypersetup{urlcolor=vividviolet}\href{https://link.springer.com/article/10.1007/BF00759302}{Gen. Rel. Grav.   \textbf{11} (1979)   391}}; 
Tanmoy Bhattacharya, Salman Habib,  Emil Mottola, ``Gravitationally induced neutrino oscillation phases in static spacetimes", {\hypersetup{urlcolor=vividviolet}\href{https://journals.aps.org/prd/abstract/10.1103/PhysRevD.59.067301}{Phys. Rev. D  \textbf{59} (1999)   067301}};
Maurizio Gasperini, ``Experimental constraints on a minimal and nonminimal violation
of the equivalence principle in the oscillations of massive neutrinos", {\hypersetup{urlcolor=vividviolet}\href{https://journals.aps.org/prd/abstract/10.1103/PhysRevD.39.3606}{Phys. Rev. D  \textbf{39} (1989)   3606}};
Nicolao Fornengo, Carlo Giunti, Chung Wook Kim, Jihye Song, ``Gravitational Effects on the Neutrino Oscillation", {\hypersetup{urlcolor=vividviolet}\href{https://journals.aps.org/prd/abstract/10.1103/PhysRevD.56.1895}{Phys. Rev. D  \textbf{56} (1997)   1895}}, \href{https://arxiv.org/abs/hep-ph/9611231}{[arXiv:hep-ph/9611231]};
Nicolao Fornengo, Carlo Giunti, Chung Wook Kim,  Jihye Song, ``Gravitational Effects on the Neutrino Oscillation in vacuum", {\hypersetup{urlcolor=vividviolet}\href{https://www.sciencedirect.com/science/article/abs/pii/S0920563298004356}{Nucl. Phys. B Proc. Suppl. \textbf{70} (1999)   264}}, \href{https://arxiv.org/abs/hep-ph/9711494}{[arXiv:hep-ph/9711494]}; Maxim Dvornikov, ``Neutrino spin oscillations in external fields in curved spacetime", {\hypersetup{urlcolor=vividviolet}\href{https://journals.aps.org/prd/abstract/10.1103/PhysRevD.99.116021}{Phys. Rev. D  \textbf{99} (2019)   116021}}, \href{https://arxiv.org/abs/1902.11285}{[arXiv:hep-ph/1902.11285]};
Luca Buoninfante, Giuseppe Gaetano Luciano, Luciano Petruzziello,  Luca Smaldone, ``Neutrino oscillations in extended theories of gravity", {\hypersetup{urlcolor=vividviolet}\href{https://journals.aps.org/prd/abstract/10.1103/PhysRevD.101.024016}{Phys. Rev. D  \textbf{101} (2020)   024016}}, \href{https://arxiv.org/abs/1906.03131}{[arXiv:gr-qc/1906.03131]};
Gaetano Lambiase, Leonardo Mastrototaro, Luca Visinelli, ``Gravitational waves and neutrino oscillations in Chern-Simons axion gravity", {\hypersetup{urlcolor=vividviolet}\href{https://iopscience.iop.org/article/10.1088/1475-7516/2023/01/011}{JCAP \textbf{01} (2023)   011}}, \href{https://arxiv.org/abs/2207.08067}{[arXiv:hep-ph/2207.08067]};
Gaetano Lambiase, Giorgio Papini, Raffaele Punzi, Gaetano Scarpetta, ``Neutrino optics and oscillations in gravitational fields", {\hypersetup{urlcolor=vividviolet}\href{https://journals.aps.org/prd/abstract/10.1103/PhysRevD.71.073011}{ Phys. Rev. D  \textbf{71} (2005)   073011}}, \href{https://arxiv.org/abs/gr-qc/0503027}{[arXiv:gr-qc/0503027]};
George Koutsoumbas, Dimitrios Metaxas, ``Neutrino oscillations in gravitational and cosmological backgrounds", {\hypersetup{urlcolor=vividviolet}\href{https://link.springer.com/article/10.1007/s10714-020-02758-z}{ Gen. Rel. Grav.   \textbf{52} (2020)   102}}, \href{https://arxiv.org/abs/1909.02735}{[arXiv:hep-ph/1909.02735]};
Maxim Dvornikov, ``Neutrino flavor oscillations in stochastic gravitational waves", {\hypersetup{urlcolor=vividviolet}\href{https://journals.aps.org/prd/abstract/10.1103/PhysRevD.100.096014}{ Phys. Rev. D  \textbf{100} (2019)  096014}}, \href{https://arxiv.org/abs/1906.06167}{[arXiv:hep-ph/1906.06167]};
Donato Bini, Christian Cherubini, Gianluca Cruciani, Andrea Lunari, ``Neutrino current in a gravitational plane wave collision background", {\hypersetup{urlcolor=vividviolet}\href{https://www.worldscientific.com/doi/abs/10.1142/S0218271803003876}{ Int. Jour. Mod. Phys. D  \textbf{12} (2003)   1983}}, \href{https://arxiv.org/abs/gr-qc/0312105}{[arXiv:gr-qc/0312105]};
Kuantay Boshkayev, Orlando Luongo, Marco Muccino, ``Neutrino oscillation in the $q$-metric", {\hypersetup{urlcolor=vividviolet}\href{https://link.springer.com/article/10.1140/epjc/s10052-020-08533-3}{ Eur. Phys. Jour. C   \textbf{80} (2020)   964}}, \href{https://arxiv.org/abs/2010.08254}{[arXiv:gr-qc/2010.08254]};
Andrea Geralico, Orlando Luongo,  ``Neutrino oscillations in the field of a rotating deformed mass", {\hypersetup{urlcolor=vividviolet}\href{https://www.sciencedirect.com/science/article/abs/pii/S0375960112002046}{  Phys. Lett. A   \textbf{376} (2012)   1239}}, \href{https://arxiv.org/abs/1202.5408}{[arXiv:gr-qc/1202.5408]};
Jun Ren, Cheng-Min Zhang,  ``Neutrino oscillations in the Kerr-Newman spacetime", {\hypersetup{urlcolor=vividviolet}\href{https://iopscience.iop.org/article/10.1088/0264-9381/27/6/065011}{  Class. Quantum Grav.   \textbf{27} (2010)    065011}}, \href{https://arxiv.org/abs/1002.0648}{[arXiv:gr-qc/1002.0648]};
Mirzabek Alloqulov, Hrishikesh Chakrabarty, Daniele Malafarina, Bobomurat Ahmedov, Ahmadjon Abdujabbarov,  ``Gravitational lensing of neutrinos in parametrized black hole spacetimes", {\hypersetup{urlcolor=vividviolet}\href{https://iopscience.iop.org/article/10.1088/1475-7516/2025/02/070}{  JCAP   \textbf{02} (2025)    070}}, \href{https://arxiv.org/abs/2408.12916}{[arXiv:gr-qc/2408.12916]}; Daniele Gregoris, ``Neutrino oscillations in the background of a black hole embedded in an expanding Universe", {\hypersetup{urlcolor=vividviolet}\href{https://www.sciencedirect.com/science/article/pii/S0550321325001671?via%3Dihub}{  Nucl. Phys. B   \textbf{1017} (2025)  116958}}; Maxim Dvornikov,  ``Spin oscillations of neutrinos scattered off a rotating black hole", {\hypersetup{urlcolor=vividviolet}\href{https://link.springer.com/article/10.1140/epjc/s10052-020-8046-z}{ Eur. Phys. Jour. C   \textbf{80} (2020)   474}}, \href{https://arxiv.org/abs/2006.01636v1}{[arXiv:hep-ph/2006.01636]}; Yuxuan Shi, Hongbo Cheng, ``The neutrino flavor oscillations in the static and spherically symmetric black-hole-like wormholes", {\hypersetup{urlcolor=vividviolet}\href{https://link.springer.com/article/10.1140/epjc/s10052-025-14460-y}{ Eur. Phys. Jour. C   \textbf{85} (2025)   909}}, \href{https://arxiv.org/abs/2412.02144v4}{[arXiv:gr-qc/2412.02144]}; Jun Ren, Yuan-Yue Pan, ``Neutrino Oscillations in the de Sitter and the Anti-de Sitter Space-Time", {\hypersetup{urlcolor=vividviolet}\href{https://link.springer.com/article/10.1007/s10773-011-0757-3}{ Int. Jour. Theo. Phys.   \textbf{50} (2011)  2614}}; Jun Ren, Yuan-Yue Pan, ``Using Hamilton-Jacobi Equation to Study the Neutrino Oscillations in the Stationary Space-Time", {\hypersetup{urlcolor=vividviolet}\href{https://link.springer.com/article/10.1007/s10773-011-0988-3}{ Int. Jour. Theo. Phys.   \textbf{51} (2012)  1111}}.
		

\bibitem{miotolman}
Donato Bini, Daniele Gregoris, Kjell Rosquist, Sauro Succi, ``Particle motion in a photon gas: friction matters", {\hypersetup{urlcolor=vividviolet}\href{https://link.springer.com/article/10.1007/s10714-012-1425-5}{Gen. Rel. Grav. \textbf{44} (2012)   2669}}.	

\bibitem{indi}
Donato Bini, Daniele Gregoris, Kjell Rosquist, Sauro Succi, ``Effects of friction forces on the motion of objects in smoothly matched interior/exterior spacetimes", {\hypersetup{urlcolor=vividviolet}\href{https://iopscience.iop.org/article/10.1088/0264-9381/30/2/025009}{Class. Quantum Grav. \textbf{30} (2013)   025009}}.

\bibitem{frolovA}
Valeri P. Frolov, ``Spinoptics in a curved spacetime", {\hypersetup{urlcolor=vividviolet}\href{https://journals.aps.org/prd/abstract/10.1103/PhysRevD.110.064020}{Phys. Rev. D  \textbf{110} (2024)  064020}}, \href{https://arxiv.org/abs/2405.01777}{[arXiv:gr-qc/2405.01777]}. 


\bibitem{frolovB}	
Valeri P. Frolov,  ``Spinoptics in the Schwarzschild spacetime", {\hypersetup{urlcolor=vividviolet}\href{https://iopscience.iop.org/article/10.1088/1475-7516/2024/12/051}{  JCAP   \textbf{12} (2024)    051}}, \href{https://arxiv.org/abs/2410.19413}{[arXiv:gr-qc/2410.19413]}.

\bibitem{frolovC}	
Valeri P. Frolov, Alex Koek,  ``Spinoptics in the Kerr Spacetime: Polarized Wave Scattering", {\hypersetup{urlcolor=vividviolet}\href{https://journals.aps.org/prd/abstract/10.1103/trh5-4sgq}{ Phys. Rev. D   \textbf{111} (2025)    104081}}, \href{https://arxiv.org/abs/2503.19208}{[arXiv:gr-qc/2503.19208]}.
	
\bibitem{tov1}
Richard C. Tolman,   ``Static Solutions of Einstein's Field Equations for Spheres of Fluid'', {\hypersetup{urlcolor=vividviolet}\href{https://journals.aps.org/pr/abstract/10.1103/PhysRev.55.364}{ Phys. Rev.  \textbf{55} (1939)  364}}.	

\bibitem{MSW}
Lincoln Wolfenstein, ``Neutrino Oscillations in Matter'',
{\hypersetup{urlcolor=vividviolet}\href{https://link.aps.org/doi/10.1103/PhysRevD.17.2369}{Phys. Rev. D  \textbf{17} (1978)  2369}}; Stanislav P. Mikheev, Alexei Yuryevich Smirnov, ``Resonant amplification of $\nu$ oscillations in matter and solar-neutrino spectroscopy'',
{\hypersetup{urlcolor=vividviolet}\href{https://link.springer.com/article/10.1007/BF02508049}{Nuovo Cim. C  \textbf{9} (1986)  17}}.
	
	
\bibitem{tolmanold}
Richard C. Tolman, Paul Ehrenfest, Boris Podolsky, ``On the Gravitational Field Produced by Light'',
{\hypersetup{urlcolor=vividviolet}\href{https://journals.aps.org/pr/abstract/10.1103/PhysRev.37.602}{Phys. Rev.   \textbf{37} (1931)  602}}.
	
	

\bibitem{ISA}
Richard A. Isaacson, ``Gravitational Radiation in the Limit of High Frequency. I. The Linear Approximation and Geometrical Optics", {\hypersetup{urlcolor=vividviolet}\href{https://journals.aps.org/pr/abstract/10.1103/PhysRev.166.1263}{Phys. Rev.  \textbf{166} (1968)   1263}}; Richard A. Isaacson, ``Gravitational Radiation in the Limit of High Frequency. II. Nonlinear Terms and the Effective Stress Tensor", {\hypersetup{urlcolor=vividviolet}\href{https://journals.aps.org/pr/abstract/10.1103/PhysRev.166.1272}{Phys. Rev.  \textbf{166} (1968)   1272}}.
	
	
\bibitem{tolmaneq}
Richard C. Tolman,  Paul Ehrenfest,   ``Temperature Equilibrium in a Static Gravitational Field",
{\hypersetup{urlcolor=vividviolet}\href{https://journals.aps.org/pr/abstract/10.1103/PhysRev.36.1791}{ Phys. Rev.   \textbf{36}  (1930)  1791}};
Richard C. Tolman,   ``Thermal Equilibrium in a General Gravitational Field",
{\hypersetup{urlcolor=vividviolet}\href{https://www.jstor.org/stable/86820?origin=JSTOR-pdf}{ PNAS    \textbf{21}  (1935)  321}}.
	
	
\bibitem{landau}
Lev D. Landau, Evgenij M. Lifshitz,  {\it Statistical Physics},  (Butterworth-Heinemann, 1980).

\bibitem{carr}
Bernard J. Carr, Alan A. Coley, ``Complete classification of spherically symmetric self-similar perfect fluid solutions", {\hypersetup{urlcolor=vividviolet}\href{https://journals.aps.org/prd/abstract/10.1103/PhysRevD.62.044023}{Phys. Rev. D  \textbf{62} (2000)   044023}}, \href{https://arxiv.org/abs/gr-qc/9901050}{[arXiv:gr-qc/9901050]}.

\bibitem{stodo}
Leo Stodolsky, ``When the Wavepacket is Unnecessary", {\hypersetup{urlcolor=vividviolet}\href{https://journals.aps.org/prd/abstract/10.1103/PhysRevD.58.036006}{Phys. Rev. D  \textbf{58} (1998)   036006}}, \href{https://arxiv.org/abs/hep-ph/9802387v2}{[arXiv:hep-ph/9802387]}.

\bibitem{bilenky2}
Samoil M. Bilenky, Serguey T. Petcov, ``Massive neutrinos and neutrino oscillations", {\hypersetup{urlcolor=vividviolet}\href{https://journals.aps.org/rmp/abstract/10.1103/RevModPhys.59.671}{ Rev. Mod. Phys.   \textbf{59} (1987)   671}}; {\hypersetup{urlcolor=vividviolet}\href{https://journals.aps.org/rmp/abstract/10.1103/RevModPhys.60.575}{ Erratum Rev. Mod. Phys.   \textbf{60} (1988)   575}}; {\hypersetup{urlcolor=vividviolet}\href{https://journals.aps.org/rmp/abstract/10.1103/RevModPhys.61.169}{ Erratum Rev. Mod. Phys.   \textbf{61} (1989)   169}}.

\bibitem{stodolsky}
Leo Stodolsky, ``Matter and light wave interferometry in gravitational fields", {\hypersetup{urlcolor=vividviolet}\href{https://link.springer.com/article/10.1007/BF00759302}{Gen. Rel. Grav.   \textbf{11} (1979)   391}}.	

\bibitem{lipkin}
Harry J. Lipkin,  ``Quantum Theory of Neutrino Oscillations for Pedestrians - Simple Answers to Confusing Questions",  {\hypersetup{urlcolor=vividviolet}\href{https://www.sciencedirect.com/science/article/abs/pii/S0370269306012299?via%3Dihub}{Phys. Lett. B \textbf{642} (2006)   366}},  \href{https://arxiv.org/abs/hep-ph/0505141}{[arXiv:hep-ph/0505141]}.	

\bibitem{referee2}
Yu-Hao Zhang, Xue-Qian Li, ``Three-generation neutrino oscillations in curved spacetime'',
{\hypersetup{urlcolor=vividviolet}\href{https://www.sciencedirect.com/science/article/pii/S0550321316302565?via%3Dihub}{Nucl. Phys. B  \textbf{911} (2016)  563}}, \href{https://arxiv.org/abs/1606.05960}{[arXiv:hep-ph/1606.05960]}.




\bibitem{NPC}
Ezra Newman, Roger Penrose, ``An Approach to Gravitational Radiation by a Method of Spin Coefficients", {\hypersetup{urlcolor=vividviolet}\href{https://aip.scitation.org/doi/10.1063/1.1724257}{Jour. Math. Phys. \textbf{3}  (1962)   566}}.  

\bibitem{szekeres}
Peter  Szekeres, ``The Gravitational Compass", {\hypersetup{urlcolor=vividviolet}\href{https://aip.scitation.org/doi/10.1063/1.1704788}{Jour. Math. Phys. \textbf{6}  (1965)   1387}}. 

\bibitem{4levels}
Bronis\l{}aw Kuchowicz,  ``Neutrinos in general relativity: Four(?) levels of approach", {\hypersetup{urlcolor=vividviolet}\href{https://link.springer.com/article/10.1007/BF00763501}{Gen. Rel. Grav. \textbf{5} (1974)   201}}.

\bibitem{ANDA}
Jeeva S. Anandan, ``Non Local Aspects of Quantum Phases'',
{\hypersetup{urlcolor=vividviolet}\href{https://www.numdam.org/item/?id=AIHPA_1988__49_3_271_0}{Ann. Inst. H. Poincar\'e Phys. Theor.   \textbf{49} (1988) 271}}.

\bibitem{berry1}
Michael V. Berry, ``Quantal Phase Factors Accompanying Adiabatic Changes", {\hypersetup{urlcolor=vividviolet}\href{https://royalsocietypublishing.org/rspa/article-abstract/392/1802/45/15579/Quantal-phase-factors-accompanying-adiabatic?redirectedFrom=fulltext}{Proc. Roy. Soc. Lond. A  \textbf{392} (1984)  45}}. 

\bibitem{berry2}
Barry Simon, ``Holonomy, the Quantum Adiabatic Theorem, and Berry's Phase", {\hypersetup{urlcolor=vividviolet}\href{https://journals.aps.org/prl/abstract/10.1103/PhysRevLett.51.2167}{Phys. Rev. Lett.  \textbf{51} (1983)  2167}}. 

\bibitem{helicity}
Myron Mathisson, ``Neue mechanik materieller systemes", {\hypersetup{urlcolor=vividviolet}\href{https://inspirehep.net/literature/48323}{Acta Phys. Polon.  {\bf 6}  (1937) 163}}; 
Achilles Papapetrou, ``Spinning Test-Particles in General Relativity. I", {\hypersetup{urlcolor=vividviolet}\href {https://www.jstor.org/stable/98893?origin=ads}{Proc. Roy. Soc. Lond. A \textbf{209}  (1951) 248}};
William Gray Dixon, ``Dynamics of extended bodies in general relativity. I. Momentum and angular momentum", {\hypersetup{urlcolor=vividviolet}\href{https://www.jstor.org/stable/2416466?origin=ads}{Proc. Roy. Soc. Lond. A \textbf{314}  (1970) 499}}.




\bibitem{boost}
Maxim Dvornikov, ``Neutrino spin oscillations in gravitational fields", {\hypersetup{urlcolor=vividviolet}\href{https://www.worldscientific.com/doi/abs/10.1142/S021827180600870X}{Int. Jour. Mod. Phys. D  \textbf{15} (2006)   1017}}, \href{https://arxiv.org/abs/hep-ph/0601095}{[arXiv:hep-ph/0601095]};


\bibitem{mash}
Bahram Mashhoon, ``Neutron interferometry in a rotating frame of reference", {\hypersetup{urlcolor=vividviolet}\href{https://journals.aps.org/prl/abstract/10.1103/PhysRevLett.61.2639}{ Phys. Rev. Lett.   \textbf{61} (1988)   2639}}.

\bibitem{Weyloriginal}
Hermann Weyl, ``Gravitation and the electron", {\hypersetup{urlcolor=vividviolet}\href{https://www.pnas.org/doi/abs/10.1073/pnas.15.4.323}{ PNAS   \textbf{15} (1929)   323}}.


\bibitem{low}
Dirk N\"otzold, Georg Raffelt, ``Neutrino dispersion at finite temperature and density", {\hypersetup{urlcolor=vividviolet}\href{https://www.sciencedirect.com/science/article/abs/pii/0550321388901137}{ Nucl. Phys. B   \textbf{307} (1988)   924}}; Juan Carlos D'Olivo, Jos\'e F. Nieves, Manuel Torres, ``Finite-temperature corrections to the effective potential of neutrinos in a medium", {\hypersetup{urlcolor=vividviolet}\href{https://journals.aps.org/prd/abstract/10.1103/PhysRevD.46.1172}{ Phys. Rev. D   \textbf{46} (1992)   1172}}.



\bibitem{high}  
G\"unter Sigl, Georg Raffelt, ``General kinetic description of relativistic mixed neutrinos", {\hypersetup{urlcolor=vividviolet}\href{https://www.sciencedirect.com/science/article/abs/pii/055032139390175O}{Nucl. Phys. B \textbf{406} (1993)  423}}; Bruce H.J. McKellar, Mark J. Thomson, ``Oscillating  neutrinos in the early universe", {\hypersetup{urlcolor=vividviolet}\href{https://journals.aps.org/prd/abstract/10.1103/PhysRevD.49.2710}{Phys. Rev. D \textbf{49} (1994)  2710}}.

\bibitem{diff}  
Eric Braaten, Robert D. Pisarski, ``Soft amplitudes in hot gauge theories: A general analysis", {\hypersetup{urlcolor=vividviolet}\href{https://www.sciencedirect.com/science/article/abs/pii/055032139090508B}{Nucl. Phys. B \textbf{337} (1990)  569}}. 

\bibitem{OHL}
Tommy Ohlsson, ``Equivalence between Gaussian averaged neutrino oscillations and neutrino decoherence'',
{\hypersetup{urlcolor=vividviolet}\href{https://www.sciencedirect.com/science/article/abs/pii/S0370269301001782?via%3Dihub}{Phys. Lett. B  \textbf{502} (2001)  159}}, \href{https://arxiv.org/abs/hep-ph/0012272}{[arXiv:hep-ph/0012272]}.


\bibitem{coh}
Carlo Giunti, Chung Wook Kim, Ung Won Lee, ``When do neutrinos really oscillate? Quantum mechanics of neutrino oscillations'',
{\hypersetup{urlcolor=vividviolet}\href{https://journals.aps.org/prd/abstract/10.1103/PhysRevD.44.3635}{Phys. Rev. D  \textbf{44} (1991)  3635}}; Carlo Giunti, Chung Wook Kim, ``Coherence of neutrino oscillations in the wave packet approach'',
{\hypersetup{urlcolor=vividviolet}\href{https://journals.aps.org/prd/abstract/10.1103/PhysRevD.58.017301}{Phys. Rev. D  \textbf{58} (1998)  017301}}, \href{https://arxiv.org/abs/hep-ph/9711363}{[arXiv:hep-ph/9711363]}; Harry J. Lipkin, ``What is coherent in neutrino oscillations'',
{\hypersetup{urlcolor=vividviolet}\href{https://www.sciencedirect.com/science/article/pii/S0370269303017325?via%3Dihub}{Phys. Lett. B  \textbf{579} (2004)  355}}, \href{https://arxiv.org/abs/hep-ph/0304187v3}{[arXiv:hep-ph/0304187]}; Am\`elie Chatelain, Maria Cristina Volpe, ``Neutrino decoherence in presence of strong gravitational fields'', {\hypersetup{urlcolor=vividviolet}\href{https://www.sciencedirect.com/science/article/pii/S037026931930872X?via%3Dihub}{Phys. Lett. B  \textbf{801} (2020)  135150}}, \href{https://arxiv.org/abs/1906.12152}{[arXiv:hep-ph/1906.12152]}; Parvaneh Sadeghi, Fay\c{c}al Hammad, Alexandre Landry, T. Martel, ``Wave packet treatment of neutrino avor oscillations in various spacetimes'', {\hypersetup{urlcolor=vividviolet}\href{https://link.springer.com/article/10.1007/s10714-021-02872-6}{Gen. Rel. Grav.  \textbf{53} (2021)  98}}, \href{https://arxiv.org/abs/2111.01441}{[arXiv:gr-qc/2111.01441]}. 

\bibitem{donpage}
Wojciech H. Zurek, Don N. Page, ``Firewalls, black-hole thermodynamics, and singular solutions of the Tolman-Oppenheimer-Volkoff equation", {\hypersetup{urlcolor=vividviolet}\href{https://journals.aps.org/prd/abstract/10.1103/PhysRevD.29.628}{Phys. Rev. D  \textbf{29} (1984)   628}}, \href{https://arxiv.org/abs/1511.07051}{[arXiv:gr-qc/1511.07051]}.

\bibitem{strong}
Erwin H. Tanin, Yikun Wang, ``Gradient-Produced Neutrinos", 
\href{https://arxiv.org/abs/2604.21968v1}{[arXiv:hep-ph/2604.21968]}.

\bibitem{GRSN}
Yue Yang, James P. Kneller, ``GR Effects in Supernova Neutrino Flavor Transformation", {\hypersetup{urlcolor=vividviolet}\href{https://journals.aps.org/prd/abstract/10.1103/PhysRevD.96.023009}{Phys. Rev. D  \textbf{96} (2017)   023009}},    
\href{https://arxiv.org/abs/1705.09723v3}{[arXiv:astro-ph.HE/1705.09723]}.

\bibitem{quiet}
Peter Goldreich, William H. Julian,  ``Pulsar Electrodynamics", {\hypersetup{urlcolor=vividviolet}\href{https://articles.adsabs.harvard.edu/pdf/1969ApJ...157..869G}{ApJ  \textbf{157} (1969)   869}}.

\bibitem{landauA}
Lev Davidovi\v{c} Landau, \lq\lq A theory of energy transfer. 2.", {\hypersetup{urlcolor=vividviolet}\href{https://www.sciencedirect.com/science/chapter/edited-volume/abs/pii/B9780080105864500146?via%3Dihub}{Physikalische Zeitschrift der Sowjetunion  \textbf{2} (1932)  46}}; Clarence Zener, \lq\lq Non-Adiabatic Crossing of Energy Levels", {\hypersetup{urlcolor=vividviolet}\href{https://royalsocietypublishing.org/rspa/article/137/833/696/3285/Non-adiabatic-crossing-of-energy-levels}{Proc. Royal Soc. of  London A  \textbf{137} (1932)  696}}; Ernst St\"ueckelberg, \lq\lq Theorie der unelastischen St\"o\ss e ùzwischen Atomen", {\hypersetup{urlcolor=vividviolet}\href{https://archive.org/details/nasa_techdoc_19720003957}{Helvetica Physica Acta \textbf{5} (1932) 369}}.	

\bibitem{wald}
Robert M. Wald,   {\it General Relativity},  (The University of Chicago Press, Chicago and London, 1984).

\bibitem{bethe}
Hans Albrecht Bethe, ``Possible Explanation of the Solar Neutrino Puzzle'',
{\hypersetup{urlcolor=vividviolet}\href{https://journals.aps.org/prl/abstract/10.1103/PhysRevLett.56.1305}{Phys. Rev. Lett.  \textbf{56} (1986)  1305}}.
	
\bibitem{MTW}
Charles W. Misner, Kip S. Thorne, John Archibald Wheeler, {\it Gravitation}, (Freeman, San Francisco, 1970).

\bibitem{entropy}  
Daniele Gregoris, Yen Chin Ong, ``Understanding gravitational entropy of black holes: A new proposal via curvature invariants", {\hypersetup{urlcolor=vividviolet}\href{https://journals.aps.org/prd/abstract/10.1103/PhysRevD.105.104017}{Phys. Rev. D \textbf{105} (2022)  104017}}, \href{https://arxiv.org/abs/2109.11968}{[arXiv:gr-qc/2109.11968]}.


\bibitem{naked1}  
Pankaj S. Joshi, Daniele Malafarina, Ramesh Narayan, ``Distinguishing black holes from naked singularities through their accretion disk properties", {\hypersetup{urlcolor=vividviolet}\href{https://iopscience.iop.org/article/10.1088/0264-9381/31/1/015002}{Class. Quantum Grav. \textbf{31} (2014)  015002}}, \href{https://arxiv.org/abs/1304.7331}{[arXiv:gr-qc/1304.7331]}; Rajibul Shaikh, Pankaj S. Joshi, ``Can we distinguish black holes from naked singularities by the images of their accretion disks?", {\hypersetup{urlcolor=vividviolet}\href{https://iopscience.iop.org/article/10.1088/1475-7516/2019/10/064}{JCAP \textbf{10} (2019)  064}}, \href{https://arxiv.org/abs/1909.10322v2}{[arXiv:gr-qc/1909.10322]}.

\bibitem{naked2}  
Avijit Chowdhury, Narayan Banerjee, ``Echoes from a singularity", {\hypersetup{urlcolor=vividviolet}\href{https://journals.aps.org/prd/abstract/10.1103/PhysRevD.102.124051}{Phys. Rev. D \textbf{102} (2020)  124051}}, \href{https://arxiv.org/abs/2006.16522v3}{[arXiv:gr-qc/2006.16522]}.

\bibitem{naked3}  
Indu K. Dihingia, Akhil Uniyal, Yosuke Mizuno, ``Distinguishability of a naked singularity from a black hole in dynamics and radiative signatures", {\hypersetup{urlcolor=vividviolet}\href{https://iopscience.iop.org/article/10.3847/1538-4357/ad9600}{Astrophys. Jour.  \textbf{978} (2025)  44}}, \href{https://arxiv.org/abs/2410.13406}{[arXiv:astro-ph.HE/2410.13406]}.

\bibitem{Thomas}  
Llewellyn Hilleth Thomas, ``The kinematics of an electron with an axis", {\hypersetup{urlcolor=vividviolet}\href{https://www.tandfonline.com/doi/abs/10.1080/14786440108564170}{Phil. Mag. Ser. \textbf{3} (1927)  1}}.



\bibitem{stul}  
Zden\v{e}k Stuchl\'ik, Jan Hlad\'ik, Jaroslav Vrba, Camilo Posada, ``Neutrino trapping in extremely compact Tolman VII spacetimes", {\hypersetup{urlcolor=vividviolet}\href{https://link.springer.com/article/10.1140/epjc/s10052-021-09309-z}{Eur. Phys. Jour. C \textbf{81} (2021)  529}}, \href{https://arxiv.org/abs/2106.05750}{[arXiv:gr-qc/2106.05750]}.


\bibitem{urca}  
Dima G. Yakovlev, Alexander D. Kaminker, Oleg Y. Gnedin, Pawel Haensel, ``Neutrino Emission from Neutron Stars", {\hypersetup{urlcolor=vividviolet}\href{https://www.sciencedirect.com/science/article/abs/pii/S0370157300001319?via%3Dihub}{Phys. Rept. \textbf{354} (2001)  1}}, \href{https://arxiv.org/abs/astro-ph/0012122}{[arXiv:astro-ph/0012122]}.

\bibitem{zeldovich}  
Ya B Zel'dovich,  Semyon Solomonovich  Gershte\v{i}n, ``Nuclear reactions in cold hydrogen. I. Mesonic catalysis", {\hypersetup{urlcolor=vividviolet}\href{https://iopscience.iop.org/article/10.1070/PU1961v003n04ABEH003311}{Sov. Phys. Usp. \textbf{3} (1961)  593}}.


\bibitem{SNA}  
Ian Padilla-Gay, Heng-Hao Chen, Sajad Abbar, Meng-Ru Wu, Zewei Xiong, ``Flavor Equilibration of Supernova Neutrinos: Exploring the Dynamics of Slow Modes", {\hypersetup{urlcolor=vividviolet}\href{https://journals.aps.org/prd/abstract/10.1103/jg14-8p4l}{Phys. Rev. D \textbf{112} (2025)  043039}}, \href{https://arxiv.org/abs/2505.11588v2}{[arXiv:astro-ph.HE/2505.11588]}.


\bibitem{supernovarep}  
Hans-Thomas Janka, Karlheinz Langanke, Andreas Marek, Gabriel Mart\'inez-Pinedo, Bernhard Mueller, ``Theory of Core-Collapse Supernovae", {\hypersetup{urlcolor=vividviolet}\href{https://www.sciencedirect.com/science/article/abs/pii/S0370157307000439?via%3Dihub}{Phys. Rept. \textbf{442} (2007)  38}}, \href{https://arxiv.org/abs/astro-ph/0612072}{[arXiv:astro-ph/0612072]}.
		
		
		
	
		
		




\end{thebibliography}
\end{document}